\documentclass{aa}

\usepackage{graphicx}
\usepackage{dcolumn}
\usepackage{bm}

\usepackage{placeins}
\usepackage{txfonts}
\usepackage{natbib}
\usepackage{amsmath}    
\usepackage{graphicx}   
\usepackage{verbatim}   
\usepackage{color}      
\usepackage{subfigure}  
\usepackage{gensymb} 
\usepackage{wasysym} 
\usepackage{hyperref}   
\bibpunct{(}{)}{;}{a}{}{,} 
\usepackage{graphicx}
\usepackage{nth}
\newcommand{\kms}{\mbox{${\rm km\,s}^{-1}$}}

\newcommand{\ms}{\mbox{${\rm m\,s}^{-1}$}}

\newcommand{\Msolar}{\mbox{${M}_{\astrosun}$}}
\newcommand{\Rsolar}{\mbox{${R}_{\astrosun}$}}

\newcommand{\Mjup}{\mbox{${M}_{J}$}}

\newcommand{\Rjup}{\mbox{${R}_{J}$}}

\newcommand\T{\rule{0pt}{2.2ex}}

\begin{document}

\title{Into the Storm: Diving into the winds of the ultra hot Jupiter WASP-76~b with HARPS and ESPRESSO}
    
\author{J.~V.~Seidel\inst{1} 
\and D.~Ehrenreich\inst{1}
\and R.~Allart\inst{2,1} 
\and H.~J.~Hoeijmakers\inst{3} 
\and C.~Lovis\inst{1}
\and V.~Bourrier\inst{1}
\and L.~Pino\inst{4} 
\and A.~Wyttenbach\inst{5} 
\and V.~Adibekyan\inst{6,7} 
\and Y.~Alibert\inst{8}
\and F.~Borsa\inst{9} 
\and N.~Casasayas-Barris\inst{13,19,20}
\and S.~Cristiani\inst{11}
\and O.~D.~S.~Demangeon\inst{6,7} 
\and P.~Di~Marcantonio\inst{10}
\and P.~Figueira\inst{12,6} 
\and J.~I.~Gonz\'{a}lez~Hern\'{a}ndez\inst{13} 
\and J.~Lillo-Box\inst{10} 
\and C.~J.~A.~P.~Martins\inst{6,14}
\and A.~Mehner \inst{12} 
\and P.~Molaro \inst{11, 18} 
\and N.~J.~Nunes\inst{15} 
\and E.~Palle\inst{13, 19}
\and F.~Pepe\inst{1}
\and N.~C.~Santos\inst{6,7}
\and S.~G.~Sousa\inst{6} 
\and A.~Sozzetti\inst{16}  
\and H.~M.~Tabernero\inst{6,17} 
\and M.~R.~Zapatero~Osorio\inst{17}
}

\institute{Observatoire astronomique de l'Universit\'e de Gen\`eve, Chemin Pegasi 51b, 1290 Versoix, Switzerland
\and Department of Physics, and Institute for Research on Exoplanets, Universit\'e de Montr\'eal, Montr\'eal, H3T 1J4, Canada
\and Lund Observatory, Box 43, S\"olvegatan 27, SE-22100 Lund, Sweden
\and INAF - Osservatorio Astrofisico di Arcetri, Largo E. Fermi 5, I-50125, Florence, Italy
\and Universit\'{e} Grenoble Alpes, CNRS, IPAG, 38000 Grenoble, France
\and Instituto de Astrof\'{\i}sica e Ci\^encias do Espa\c co, CAUP, Universidade do Porto, Rua das Estrelas, 4150-762 Porto, Portugal
\and Departamento de F\'isica e Astronomia, Faculdade de Ci\^encias, Universidade do Porto, Rua do Campo Alegre, 4169-007 Porto, Portugal
\and Physikalisches Institut \& NCCR PlanetS, Universit\"{a}t Bern, CH-3012 Bern, Switzerland
\and INAF -- Osservatorio Astronomico di Brera, Via E. Bianchi 46, 23807 Merate (LC), Italy
\and Centro de Astrobiolog\'{i}a (CAB, CSIC-INTA), Departamento de Astrof\'{i}sica, ESAC campus 28692 Villanueva de la Ca\~{n}ada, Madrid, Spain
\and  INAF- Osservatorio Astronomico di Trieste, via Tiepolo 11, I-34143 Trieste, Italy
\and European Southern Observatory (ESO) - Alonso de Cordova 3107, Vitacura, Santiago, Chile
\and Instituto de Astrof\'{i}sica de Canarias, Via Lactea sn, 38200, La Laguna, Tenerife, Spain
\and Centro de Astrof\'{\i}sica, Universidade do Porto, Rua das Estrelas, 4150-762 Porto, Portugal
\and Instituto de Astrof\'{i}sica e Ci\^{e}ncias do Espa\c{c}o, Faculdade de Ci\^{e}ncias da Universidade de Lisboa, Campo Grande, PT1749-016 Lisboa, Portugal
\and INAF - Osservatorio Astrofisico di Torino, Strada Osservatorio, 20 I-10025 Pino Torinese (TO), Italy
\and Centro de Astrobiolog\'{\i}a (CSIC-INTA), Ctra. de Ajalvir km 4, 28850, Torrej\'{o}n de Ardoz, Madrid, Spain
\and Institute for Fundamental Physics (IFPU), Via Beirut 2, 34151 Grignano TS, Italy
\and Departamento de Astrof\'{i}sica, Universidad de La Laguna, 38200 San Cristobal de La Laguna, Spain
\and Leiden Observatory, Leiden University, Postbus 9513, 2300 RA, Leiden, The Netherlands
}

\date{Received date/ Accepted date}

\abstract{\textit{Context.} Despite swift progress in the characterisation of exoplanet atmospheres in composition and structure, the study of atmospheric dynamics has not progressed at the same speed. While theoretical models have been developed to describe the lower layers of the atmosphere and, disconnected, the exosphere, little is known about the intermediate layers up to the thermosphere.\\ 
\textit{Aims.} We aim to provide a clearer picture of atmospheric dynamics for the class of ultra hot Jupiters, highly-irradiated gas giants, on the example of WASP-76~b. \\ 
\textit{Methods.} We analysed two datasets jointly, obtained with the HARPS and ESPRESSO spectrographs, to interpret the resolved planetary sodium doublet. We then applied the MERC code, which retrieves wind patterns, speeds, and temperature profiles on the line shape of the sodium doublet. An updated version of MERC, with added planetary rotation, also provides the possibility to model the latitude dependence of the wind patterns.\\ 
\textit{Results.} We retrieve the highest Bayesian evidence for an isothermal atmosphere, interpreted as a mean temperature of $3389\pm227$ K, a uniform day-to-night side wind of $5.5^{+1.4}_{-2.0}\, \kms$ in the lower atmosphere with a vertical wind in the upper atmosphere of $22.7^{+4.9}_{-4.1}\, \kms$, switching atmospheric wind patterns at $10^{-3}$ bar above the reference surface pressure ($10$ bar).\\ 
\textit{Conclusions.}  Our results for WASP-76~b are compatible with previous studies of the lower atmospheric dynamics of WASP-76~b and other ultra hot Jupiters. They highlight the need for vertical winds in the intermediate atmosphere above the layers probed by global circulation model studies to explain the line broadening of the sodium doublet in this planet. This work demonstrates the capability of exploiting the resolved spectral line shapes to observationally constrain possible wind patterns in exoplanet atmospheres, an invaluable input to more sophisticated 3D atmospheric models in the future.
}


\maketitle

\section{Introduction}

The study of hot Jupiters is revealing a more and more detailed picture of their atmospheric composition. Thanks to transmission spectroscopy, the existence of a wide range of atoms, ions and molecules in hot Jupiter atmospheres has been inferred \citep[e.g. ][]{Evans2018,CasasayasBarris2019, Hoeijmakers2019, Hoeijmakers2020, Tabernero2020}, and established results were revised with higher resolution \citep{CasasayasBarris2021}.
We can also detect the presence of clouds and hazes \citep[e.g. ][]{Powell2019, Parmentier2020, Gao2020, Barstow2020, Allart2020}, the thermal structure of the atmosphere \citep[e.g. ][]{Line2016, Evans2018, Gibson2020, Yan2020,Baxter2020}, even chemical abundances of the detected elements \citep[e.g. ][]{Line2014, Brogi2019, Pino2020}, and condensation of iron \citep{Ehrenreich2020, Borsa2021}. 

\noindent The first attempt to measure atmospheric winds directly, based on prior theoretical work \citep{Knutson2008}, was performed with the CO band of HD209458~b, and found a blueshift of 2~\kms compared to the systemic velocity, indicating a day-to-night side wind \citep{Snellen2010}. Similarly, studying the bands of CO and H$_2$O in the exoplanet $\beta$ Pic b in \cite{Snellen2014}, they found a rotational velocity of $v_{\mathrm{rot}}=25\pm3 \kms$. The conceptual proof of this technique started further studies constraining the rotational speed in the lower atmosphere of hot Jupiters, most of them consistent with the tidally-locked nature of these planets \citep{Louden2015,Brogi2016}.

The sodium doublet is especially well-suited to bridge the missing pressure regions between the lower atmosphere probed with molecular bands and the exosphere. It is a resonant doublet and probes the terminator up until the thermosphere in high-resolution \citep{Wyttenbach2015}. Based on this result, the HEARTS (Hot Exoplanet Atmospheres Resolved with Transmission Spectroscopy) survey was created and has provided a number of interesting datasets for the study of atmospheric winds in high-resolution \citep{Wyttenbach2017, Seidel2019, Hoeijmakers2020, Seidel2020b, Seidel2020c}. With the possibility of resolving the line shape of single spectral lines from high-resolution transmission spectra, vertical layers below the thermosphere have become accessible, via the sodium doublet and similar lines.
\noindent For example, the Balmer lines have recently been used by \cite{Wyttenbach2020} to study Parker winds and rotational broadening in the ultra hot Jupiter KELT-9~b, and by \cite{Cauley2020} to study the rotational velocity of the ultra hot Jupiter WASP-33~b. While \cite{Wyttenbach2020} find zonal wind speeds consistent with the planetary rotation for KELT-9~b, \cite{Cauley2020} find evidence of large zonal wind speeds across the terminator and tentative evidence of a day-to-night side wind. In \cite{Seidel2020}, a nested sampling algorithm is combined with a forward model of different wind patterns to retrieve the line shape of the sodium doublet for HD189733~b, thus exploring the full parameters space created by the wind speed, direction, temperature and continuum level. The retrieval revealed a high velocity outwards expanding wind in the intermediate layers of the atmosphere that drives the atmospheric mass-loss in the exosphere.

Probing the lower atmosphere by utilising the iron signature in WASP-76~b, \cite{Ehrenreich2020} measure a combination of planetary rotation and day-to-night side winds with ESPRESSO. Revisiting the data on WASP-76~b taken with HARPS, \cite{Kesseli2021} were able to confirm the results with the same technique even on the less precise HARPS data, opening up an avenue to revisit HARPS data for a multiude of similar exoplanets. The sodium doublet of WASP-76~b, first detected in \cite{Seidel2019}, was observed with ESPRESSO recently which provided an unparalleled precision of the line shape with a single transit \citep{Tabernero2020}. Based on this work, we study the atmospheric winds in WASP-76~b via the sodium doublet, probing the atmosphere up to the thermosphere, with the MERC retrieval code \citep{Seidel2020}.

\section{WASP-76~b data sets}
\label{sec:dataset}

We used two different datasets for WASP-76~b in this study, taken with the HARPS and ESPRESSO spectrographs. The HARPS WASP-76~b sodium doublet data published in \cite{Seidel2019} is based on three transits taken as part of the HEARTS survey (ESO programme: 100.C-0750; PI: Ehrenreich). However, since then, the system parameters for WASP-76 were updated. We address the impact of these changes in the following section.

\subsection{Revised system parameters and impact on HARPS dataset}

Since the analysis of the sodium doublet from the HARPS data for WASP-76~b \citep{Seidel2019}, a close companion to the host star was detected by \citet{Bohn2020} with a separation of $0.436\pm0.003\,$~arcsec and K-band magnitude difference of $2.30\pm0.05$. In consequence, the planetary parameters were updated in \citet{Southworth2020} and \citet{Ehrenreich2020}. This analysis uses the new parameters from Table \ref{tab:para} with the stellar companion, instead of the values used in \citet{Seidel2019} from the discovery paper \citet{West2016}. The change in parameters can influence the relative velocities of the system, the star, and the planet and thus lead to a different line shape and we subsequently decided to re-analyse the HARPS dataset. The two velocities used in building the transmission spectrum and which can influence the line shape are the semi-amplitude of the stellar radical velocities (RVs), $K_{\star}$, and the systemic velocity, $v_{\mathrm{sys}}$,  which are both taken from \citet{Ehrenreich2020}, see Table \ref{tab:para}. The parameters were verified in \citet{Tabernero2020} where a 2D map of the sodium detection was created for the here used dataset, clearly showing the planet trace in the stellar rest frame \citep[][Figure 5]{Tabernero2020}. An analysis of the impact of the planetary orbital velocity on the here presented results is provided in Appendix \ref{app:kpvsys}, which shows that the broadening of the lines cannot stem from imprecise values of the orbital parameters.

\begin{table}[ht]
\centering   
\caption{\label{tab:para}Planetary and stellar parameters}
\begin{tabular}{ll} \hline \hline                                          
\multicolumn{2}{l}{Taken from \cite{Ehrenreich2020}:} \T                                                        \\
\hline    
Planetary radius, $ R_{p} $ [{\Rjup}]            \T & $ 1.854  _{-0.076}^{+0.077}               $ \\                                         
Planetary mass, $ M_{p} $ [{\Mjup}]           \T & $ 0.894   _{-0.013}^{+0.014}                 $ \\
Stellar radius, $R_\ast$  [{\Rsolar}]             \T & $ 1.756 \pm 0.071               $ \\ 
Stellar mass, $ M_{\ast} $ [{\Msolar}]           \T & $ 1.458 \pm 0.021                $ \\
Orbital semi-major axis, $ a $ [au]              \T & $ 0.0330 \pm 0.0002          $ \\
Scaled $a$, $ a/R_{\ast} $                                   \T & $ 4.08   _{-0.02}^{+0.06}                 $ \\
Inclination [deg]                                \T & $ 89.623  _{-0.034}^{+0.005}                 $ \\
Eccentricity, $ e $ (fixed)                      \T & $ 0.0                                     $ \\
Transit depth, $ \Delta T$ \T                    \T & $0.01178 _{-0.00076}^{+0.00077}           $ \\
 $K_{\star}$ [\ms]                                \T & $ 116.02 _{-1.35}^{1.29}             $ \\
  $v_{\mathrm{sys}}$ [\kms]                                \T & $ -1.11 \pm 0.50            $ \\
Mid-transit time,[BJD] - 2450000                \T & $  8080.626165 _{-0.000367}^{+0.000418}                    $ \\
Transit duration [d]                            \T & $  0.15972       $ \\
Ingress duration [d]                            \T & $  0.01639           $ \\
Period [d]                                      \T & $  1.80988198 _{-0.00000056}^{+0.00000064} $ \\
Impact parameter                              \T & $  0.027  _{-0.023}^{+0.13}                 $ \\
$T_{eq}$ [K]	           \T & $ 2228  \pm 122                $ \\
 \hline
\end{tabular}
\end{table}

\begin{figure*}[htb]
\resizebox{\textwidth}{!}{\includegraphics[trim=2.0cm 0.0cm 2.0cm 1.0cm]{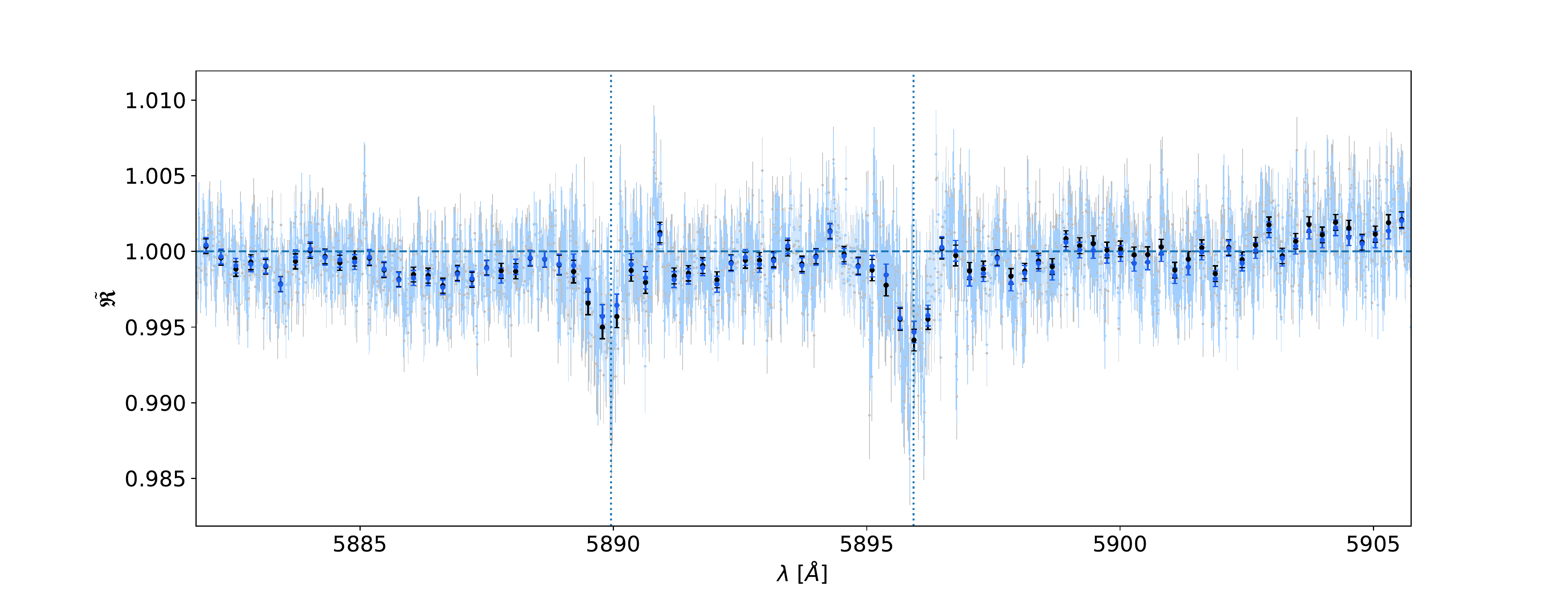}}
	\caption{HARPS datasets before and after the system parameter update normalised to 1. In black, the HARPS dataset as published in \cite{Seidel2019} and in blue the re-analysed dataset with the new system parameters. The vertical, dashed blue lines indicate the line centre of the sodium doublet lines.}
	\label{fig:HARPSspec}
\end{figure*}

\begin{figure*}[htb]
\resizebox{\textwidth}{!}{\includegraphics[trim=0.0cm 0.0cm 0.0cm 0.0cm]{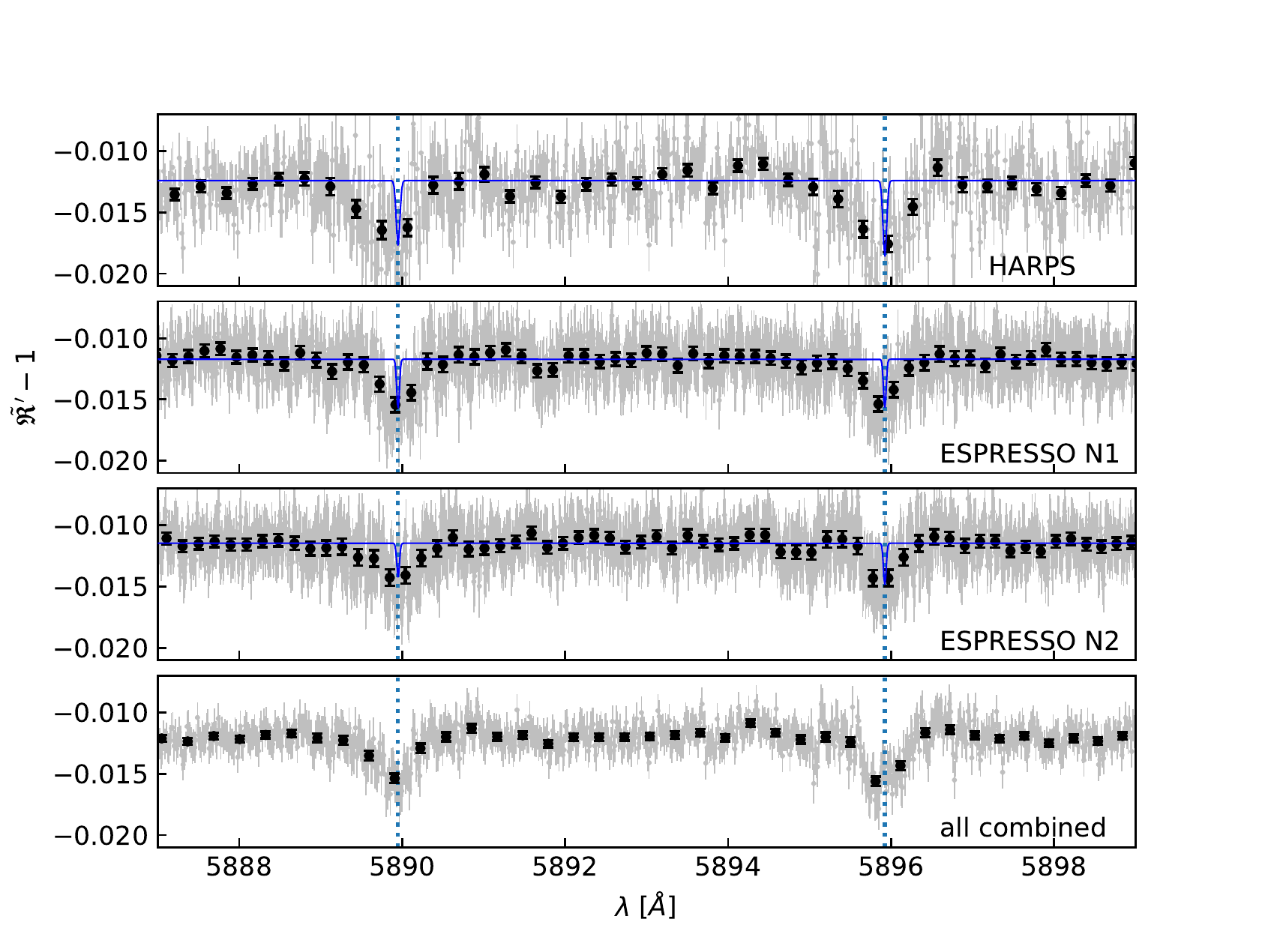}} 
	\caption{Overview of the sodium doublet transmission spectra used in this work, normalised to 0 with added contribution of the planet's obscure disk. The disk contribution depends on the used white light radius and is therefore only a best-effort approximation for the real continuum in the order of the sodium doublet. Since the same radius is also used for the retrieval, this has no impact on our conclusions. In the top panel the HARPS spectrum from \cite{Seidel2019} reprocessed with the updated system parameters, in the centre panels the ESPRESSO spectrum for each night and in the bottom panel all three datasets combined. The line centre of the sodium lines are indicated as dashed, blue, vertical lines, and the line spread function of the respective instrument as a blue graph.}
	\label{fig:data_overview}
\end{figure*}

\noindent We reduce the data in the same way as \cite{Seidel2019} but with the updated system parameters. We use the 2D echelle spectral images produced by the HARPS Data Reduction Pipeline (DRS v3.5) and discard 15 exposures in night three affected by cirrus clouds just as in \cite{Seidel2019}. We then applied {\tt molecfit} \citep{Smette2015, Kausch2015, Allart2017} to correct for tellurics and shift all spectra to the stellar rest frame. We combine all out-of-transit spectra to the normalised master-out and extract the planetary signal by dividing each in-transit spectra with the master-out. Lastly, we create the transmission spectrum by shifting the in-transit spectra into the planetary rest frame and sum over them for a sufficient S/N to detect the sodium feature.
Figure \ref{fig:HARPSspec} shows the comparison between the spectrum as presented in \cite{Seidel2019} and the new analysis with the updated parameters. The difference between the two datasets is mainly driven by the updated values for the radii. The new dataset is also shown in the upper panel of Figure \ref{fig:data_overview}. While the changes are less than $1\%$ compared to the spectrum published in \cite{Seidel2019}, both lines of the sodium doublet are less deep and narrower than before, which justifies the new reduction of the data with regards to the aim of this work: to extract information from the sodium doublet's line shape. However, the impact of the updated system parameters does not change any of the conclusions drawn in \cite{Seidel2019}.

\subsection{ESPRESSO dataset}

The transmission spectrum of WASP-76~b around the sodium doublet was created from observations with the ESPRESSO spectrograph at ESO's VLT telescope in Paranal, Chile \citep{Pepe2021}. The observations were performed as part of ESO program 1102.C-0744 (ESPRESSO GTO, \citealt{Ehrenreich2020, Tabernero2020}). Two transits of WASP-76~b were recorded, on the 03-Sept-2018 (night one) and the 31-Oct-2018 (night two), for an overview of observational conditions, see \cite{Tabernero2020}. In night one, four spectra are removed due to high airmass\footnote{airmass $<2.0$, for consistency with HARPS analysis, ESPRESSO ADC restrictions at $2.2$.}, in night two three spectra are discarded likewise.

\noindent To have planetary spectra obtained with the same technique (thus minimizing the systematics that different methods of spectra extraction might introduce in our analysis), we re-reduced the ESPRESSO data. The trace of the stellar sodium line centre is masked to eliminate residuals in all spectra with a window of $\pm4\, \kms$. Each remaining spectrum is corrected for cosmics and the transmission spectrum built following \cite{Seidel2019} and \cite{Seidel2020b}. We apply {\tt molecfit} \citep{Smette2015, Kausch2015, Allart2017} to correct both nights for the influence of telluric lines and use the SKYSUB S2D output files of the DRS pipeline, which are corrected for the impact of sky emission and blaze corrected. We corrected the telluric lines down to the noise level for all airmasses and verified that no residuals are visible in the combined master spectrum. As shown in \citet{Seidel2019} for the HARPS data and \cite{Tabernero2020} for the ESPRESSO data, the Rossiter-McLaughlin effect and centre-to-limb effects induce variations in the transmission spectra at the level of $0.04~\%$, smaller than the combined error on our final dataset $\pm0.1~\%$. Two possible techniques to correct the RM effect, a numerical approximation \citep{Wyttenbach2020} or modelling \citep[e.g.][]{CasasayasBarris2019} were compared in \citet{Seidel2020} for the hot Jupiter HD189733~b and showed differences of approximately $0.1~\%$, highlighting that the selection of different correction techniques introduces different systematic errors on the line shape. As shown in \citep{CasasayasBarris2021}, the RM correction also depends on the synthetic stellar spectra used and more work is needed to properly establish which model spectra should be used for RM corrections. However, for the ESPRESSO dataset, which is more sensitive to the RM effect, we mask the trace of the stellar sodium line centre, which coincides with area of maximum impact from the RM effect. In conclusion, for the particular dataset presented here a correction of the RM effect would introduce a larger uncertainty on our dataset and we opted to not correct for this effect in the following, but to minimise its impact through masking.

\noindent Because of interference patterns created by Coude Train optics in ESPRESSO, the combined transmission spectrum shows sinusoidal noise (wiggles) instead of a straight base line \citep{Allart2020,Tabernero2020}. A sinusoidal fit is performed on the final transmission spectrum for each night to correct for this feature. We combine both nights and both orders containing the sodium doublet ($116$ and $117$) and show the resulting final transmission spectrum in the middle panel of Figure \ref{fig:data_overview}. The independent re-analysis of the ESPRESSO transmission spectrum is compatible with the spectrum presented in \cite{Tabernero2020}.

\subsection{Combined dataset and compatibility}
\label{sec:combined}

\begin{table}
\caption{Relative absorption depth atmospheric sodium on WASP-76~b for each of the nights separately, both lines combined.}
\label{table:absorbnightly}
\centering
\begin{tabular}{l c c c }
\hline
\hline
dataset& abs. depth (\%) & reference \\
\hline
HARPS N1& $0.419\pm0.078$& \cite{Seidel2019}\\
HARPS N2& $0.361\pm0.060$&\cite{Seidel2019}\\
HARPS N3&$0.362\pm0.045$&\cite{Seidel2019}\\
\hline
HARPS combined& $0.381\pm0.036$&\\
\hline
ESPRESSO N1& $0.417\pm0.035$&\cite{Tabernero2020}\\
ESPRESSO N2&$0.270\pm0.028$&\cite{Tabernero2020}\\
\hline
\end{tabular}
\end{table}

Both the combined HARPS dataset and the ESPRESSO datasets for both nights can be found in Figure \ref{fig:data_overview}. ESPRESSO has a higher throughput and is installed on a larger telescope than HARPS. Therefore, depending on the sky conditions, it is expected that 3-4 HARPS transits, as in our HARPS dataset, are equivalent to one ESPRESSO transit. While the line shapes are roughly reproduced in all three datasets, the datasets show a difference in line depths. As already analysed in \cite{Seidel2019} for the HARPS dataset, the three combined HARPS nights were affected by thin cloud cover, and thus varying line depths, that had to be corrected. While the line depths within the HARPS dataset were subsequently compatible, the uncertainty of HARPS is comparatively large. The two nights of ESPRESSO data re-analysed here were first studied in \cite{Tabernero2020} who found an average line depth of $0.417\pm0.035~\%$ in night one and $0.270\pm0.028~\%$ in night two (see overview in Table \ref{table:absorbnightly}). In consequence, the three datasets show no visible variation in line shape when rescaled beyond what is expected from uncertainties, but differ in line depth. Stellar activity was monitored for the HARPS dataset with simultaneous photometric observations \citep{Seidel2019} and for the ESPRESSO dataset via the \ion{Ca}{ii} H and K lines, again showing no stellar activity that could account for the differing line depths. Alternatively, the difference could stem from a change in mean atmospheric temperature between the nights and thus a change in transit depth. Therefore, the value retrieved for the mean temperature of the atmosphere should be seen as a mean value and upper boundary for the real temperature of the atmosphere (see Section \ref{sec:results}). Additionally, to reduce the impact of temporal temperature variations, the three datasets were combined to obtain a more precise dataset: 

\noindent Given that the doublet lines are much broader than the line spread functions of the instruments in question (12, 11, and 13 times wider from top to bottom in Figure \ref{fig:data_overview}), we can re-bin the ESPRESSO data ($R\sim 138,000$) on the lower resolution HARPS wavelength grid ($R\sim 115,000$) and combine the three datasets of similar resolution (combined HARPS data, ESPRESSO night one, and night two). We weight each dataset with the respective mean S/N over the order of the sodium doublet: $ 48.58$, $61.20$ and $53.39$ respectively, corresponding to a fraction of $29.77\%$, $37.51\%$, and $32.72\%$ in the combined transmission spectrum. The resulting combined transmission spectrum of the sodium doublet, which is comprised of the three equivalent datasets and used in the rest of this work, is shown in the lower panel of Figure \ref{fig:data_overview}. 


\noindent A possible caveat for the study of line shapes is Doppler-smearing, introduced by the continuous change of the planet orbital velocity during one exposure \citep{RiddenHarper2016, Wyttenbach2020, Cauley2020} and the line spread function of the instrument. The exposure times for the HARPS datasets and ESPRESSSO night two were on average 300~s, for ESPRESSO night one the exposure time was 600~s. The instrument and telescope overheads, and thus the time between exposures is 32~s for HARPS and 68~s for ESPRESSO. We estimate the maximum Doppler-smear of the planet orbital velocity from the exposure times at $2.4\, \kms$ for the HARPS and night two ESPRESSO dataset, and $4.7\, \kms$ for the night one ESPRESSO dataset. This means that the mean smear over the entire exposure is approximately one resolution element and has little impact on the data for this particular system, however, we address the additional uncertainty in the discussion. MERC accounts for the instrument resolution directly.  


\section{Updates to MERC}
\label{sec:MERC}

\noindent MERC \citep{Seidel2020} combines a nested-sampling retrieval algorithm with a forward model to produce theoretical line profiles and then compare them with the provided dataset. The forward model includes Doppler-broadening by a variety of global wind patterns, which are described below. The algorithm provides the Bayesian evidence $|\ln\mathcal{Z}|$ for each forward model and the median of the marginalised posterior distributions of the parameters, the best-fit parameters, within the set parameter space. The difference in Bayesian evidence $|\ln\mathcal{B}_{01}|$ then ranks the models between each other and shows the significance of the selection of one model over another via the Jeffrey's scale (see Table \ref{table:Bayesianoverview} or \citealt{Trotta2008} and \citealt{Skilling2006}). Further details on MERC can be found in \cite{Seidel2020}.

\begin{table}[bh]
\caption{Empirical scale to judge the evidence when comparing two models $M_0$ and $M_1$, called a 'Jeffrey's scale'. The following table together with a more in depth explanation can also be found in \cite{Lavie2016} or \cite{Seidel2020}.}
\label{table:Bayesianoverview}
\centering
\begin{tabular}{c c c l }
\hline
\hline
$|\ln\mathcal{B}_{01}|$   & Odds & Probability & Strength of evidence     \\
\hline
$<1.0$  &  $<3:1$   &  $<0.750$  &   Inconclusive \\
$1.0$  &  $\sim 3:1$   &  $0.750$  &  Weak evidence \\
$2.5$  &  $\sim 12:1$   &  $0.923$  &   Moderate evidence \\
$5.0$  &  $\sim 150:1$   &  $0.993$  &   Strong evidence \\
   \hline
\end{tabular}
\end{table}

 \subsection{Wind patterns}
\label{sec:wind}

\begin{figure*}[htb]
\resizebox{\textwidth}{!}{\includegraphics[trim=-2.0cm 2.0cm -2.0cm 2.0cm]{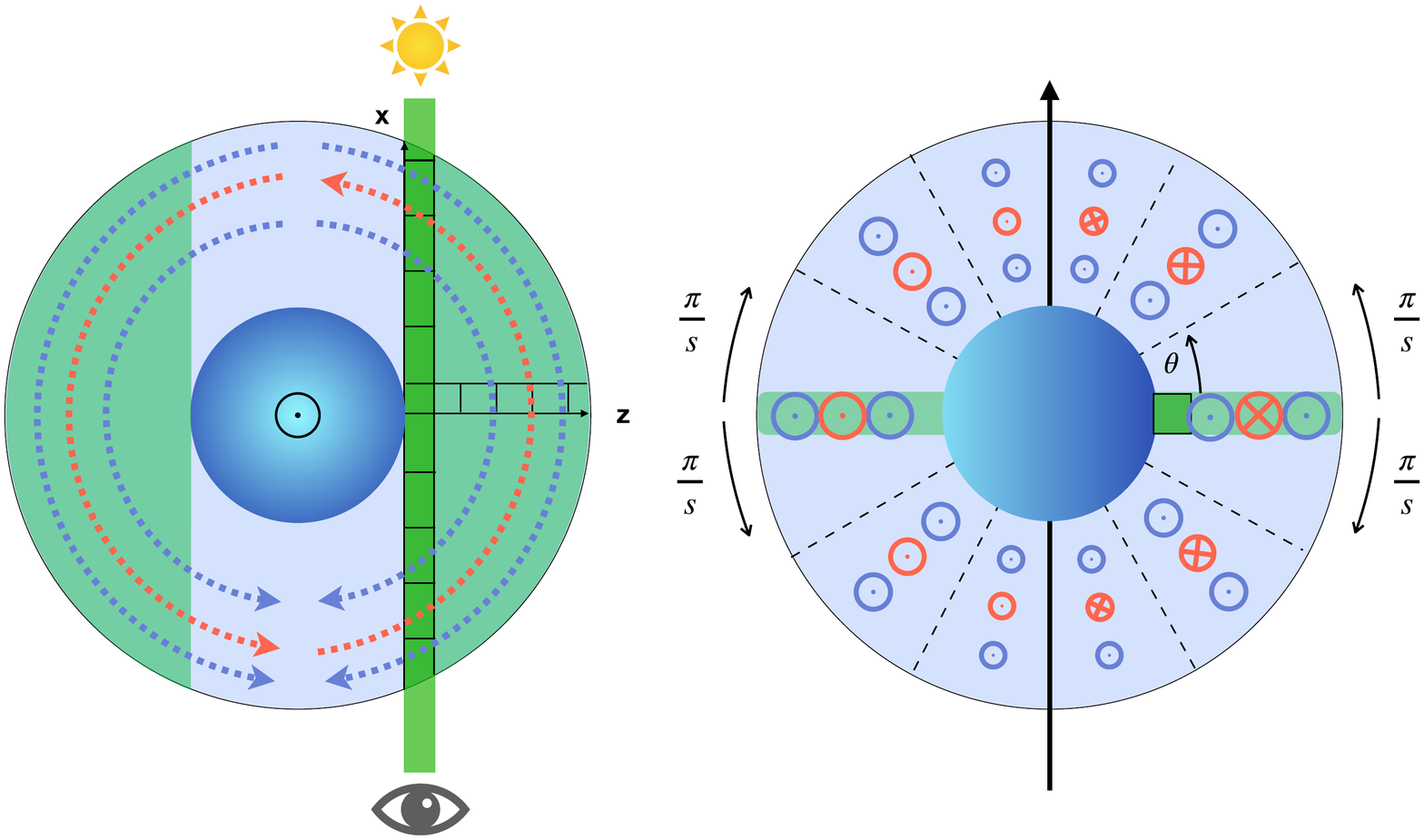}}
	\caption{Illustration of the implementation of the day-to-night side wind (updated version of Figure $3$ from \cite{Seidel2020}). All other zonal wind patterns are implemented in the same fashion. A polar view of the atmosphere is shown on the left and an equatorial view on the right. The wind direction and strength is indicated in blue, the direction and strength of the rotational velocity of the planet in red. Dashed, black lines indicate the border between sectors. For illustration purposes, we set $s = 3$, see text. \textbf{Left:} The dotted arrows indicate the direction of the velocities. The extinction coefficient is calculated in altitude along the z axis and then transposed in the x direction along the LOS (dark green cells). The LOS is then iterated upward in z until the top of the atmosphere is reached and all values are saved in a 2D grid (here visualised as a slice in light green). In each bin of the 2D grid, the combined velocity of wind and planet rotation, and the broadened profile are calculated and stored (see \cite{Seidel2020} for more details). \textbf{Right:} The reader is in the position of the observer. Points indicate a flow towards the reader, crosses away from the reader. The process described on the left is repeated for each sector, where the velocities are adjusted for the latitude via Equation \ref{eq:windlat} and then rotated by $\pi/s$ (or $\pi/2s$, depending on the symmetry) to create the full atmosphere. In this simplification of the atmosphere, the wind is not parallel to the equator, but point towards the anti-stellar point. This reduces calculation time significantly, given that it reduces the problem from 3D to 2D for each sector, with the extinction coefficient only calculated once in 1D, while providing quasi-3D insights into the wind structure.}
	\label{fig:dtnlatillustration}
\end{figure*}

MERC can model different wind patterns impacting the spectral line shape and offset: a day-to-night-side wind, a super-rotational wind and a vertical wind, as well as a combination of the first two patterns in the lower atmosphere and a vertical wind in the upper atmosphere.

\noindent To implement these wind patterns in the line of sight (LOS), we calculate the wind broadening for one atmospheric slice in the LOS (see \cite{Seidel2020} for details or Figure \ref{fig:dtnlatillustration} for an illustration). The atmospheric slice is divided into cells and each atmospheric profile in the respective cells is Doppler-shifted by the corresponding wind velocity component in the LOS. Summing over the cells to create the slice then leads to an overall Doppler-broadening. For a full 3D treatment of the atmospheric dynamics, this procedure would have to be repeated for each cell the light passes through, which requires computing times incompatible with a nested-sampling retrieval approach. Therefore, \cite{Seidel2020} uses the symmetry of the proposed wind patterns and rotates the atmospheric slice by $2\pi$ or a fraction thereof to obtain the full atmosphere. While preferable for computational reasons, this approach restricts the wind patterns to constant wind speeds in latitude throughout the planet atmosphere. Additionally, co-rotation of the atmosphere with the planet's rotation can not be accounted for, due to its latitude dependence. Especially for a super-rotational (srot) or day-to-night side (dtn) wind, work with global circulation models (GCMs) has shown that they have a more jet-like structure for hot Jupiters, with stronger winds at the equator than at the poles \citep[e.g. ][]{Showman2009}. 
 
 Subsequently, we update MERC and introduce the possibility for a solid-body like latitude dependency of the super-rotational or day-to-night side wind speeds, as was implemented for Parker winds and rotation in \cite{Wyttenbach2020}. The vertical wind only expands the atmosphere, irrespective of the latitude. This approach produces the strongest wind at the equator and imposes a reduction of zonal winds to zero at the poles. The new calculation of the wind speed in the LOS, compared to Eq. $7$ from \cite{Seidel2020} is then

\begin{equation}
|\vec{v}|_{\mathrm{LOS}}=\pm|\vec{v}_{\mathrm{srot/dtn}}|\cdot \frac{R_p+b}{\sqrt{(R_p+b)^2+x^2}}\cdot \cos \theta,
\label{eq:windlat}
\end{equation}

where $\theta$ is the angle between the equator and the radial vector to the current cell (an approximation of its latitude), $R_p$ the planet's white light radius, $b$ the impact factor (the current position above the surface in z, see Figure 2 in \cite{Seidel2020} for a visualisation), and $x$ the current position of the cell along the LOS, expressing the position of the cell in cartesian coordinates. In this treatment of atmospheric slices, the zonal winds are not parallel to the equator, but point towards the anti-stellar point (see Figure \ref{fig:dtnlatillustration}). This geometry has no impact on the LOS component of the velocity, which drives the line broadening and is only an artefact of the chosen geometry.

The planetary rotation, assuming the planet is tidally locked, adds a velocity of

\begin{equation}
|\vec{v}_{\mathrm{planet, rot}}|_{\mathrm{LOS}}= \frac{2\pi (R_p+b)}{P} \cdot \cos \theta,
\label{eq:planetvel}
\end{equation}

to each cell with $P$ as the rotation period. The planetary rotation is added to all models with or without winds.

The two velocity grids, from the planetary rotation and the wind are co-added to obtain the LOS velocity. Given that the stored atmospheric slice has to be broadened by the LOS wind speed in each cell, this computation cannot be made for each possible angle $\theta$. We introduce a sector variable $s$, that slices each quarter of each hemisphere (see Figure \ref{fig:dtnlatillustration} or \citealt{Wyttenbach2020}) into sub-sectors. The velocities are, therefore, only calculated once for each sector and then rotated by $\frac{\pi}{s}$. $\theta$ is defined as the angle in the centre of the sector. We tested different numbers of sectors (between 3 and 48 in steps of 3) and found changes of less than $1\%$ in the final model spectrum for $s\geq 9$. In consequence, all models in this work are calculated with nine sectors. The approach of introducing latitude dependent sectors introduces a new dimension to MERC and thus a quasi 3D calculation of the atmosphere at the terminator probed by transmission spectroscopy. However, some assumptions are still made for the sake of reducing the computational cost: We assume that the planet is tidally locked and that the planetary obliquity is perpendicular to the orbital plane. Both these assumptions are reasonable for exoplanets as close to their host star as WASP-76~b \citep{West2016}. 

\noindent Additionally, as already mentioned in \cite{Seidel2020}, the zonal wind patterns, now including the planetary rotation, are not axisymmetric with the rotational axis of the planet, but spherically symmetric. Since we integrate over the entire terminator and are only interested in the LOS components, this has no significant impact on the resulting line profile. On a similar note, for proper time-resolved treatment of the day-to-night side wind, we would have to take into account the orbital inclination and angle of the current position ($\zeta$ in Figure 3 of \cite{Ehrenreich2020}), given that the real symmetry axis is the connector between star and planet in a tidally locked system and not the LOS. The $\cos(i)$ of the orbital inclination is close to $1$ and subsequently ignored and the orbital angle can be disregarded as well, given that we integrate over the entire transit, averaging at transit centre where the real symmetry axis and the LOS are parallel.

\noindent We apply both the latitude dependent wind patterns as well as the unmodulated, constant wind patterns to see if winds in WASP-76~b show a tendency to be restricted into zonal jets or if a more uniform wind flow along the terminator is preferred. We take the planetary rotation into account for all retrievals in this work.

\subsection{Degeneracies and the continuum}
\label{sec:degen}

We have not modified the approach in \cite{Seidel2020} regarding the fit to the spectral continuum, where we follow the simple description of the degeneracy between the sodium abundance and the pressure scaling as presented in \cite{Lecavelier2008} and \cite{Heng2015} and retrieve a degeneracy parameter NaX and the temperature T to set the continuum level of the flux. While this is an important subject to address for exoplanets with richer chemistry and an interest in the correct retrieval of varying abundances, we fix the abundance to a constant value, which is a common approach in modelling hot Jupiter atmospheres \citep{Lecavelier2008, Agundez2014, Steinrueck2019}. Nonetheless, the impact of varying sodium abundance on the results is discussed qualitatively in Section \ref{sec:results}. With respect to the temperature, we explore a deviation from an isothermal treatment of the atmosphere by introducing a temperature gradient to verify if the common assumption of an isothermal atmosphere can also be applied here. Lastly, in \cite{Seidel2020}, MERC retrieved the best-fit parameters in two stages, first on the continuum, far from the line doublet, to get the appropriate range of possible parameters for $NaX$ and $T$, and then on the line to further constrain the continuum parameters and the wind speeds and directions. The update to the code presented here includes additional parallelisation and now allows to select a slightly wider wavelength range of three times the FWHM, thus combining the line core with enough of the continuum to eliminate the need for a two-tier retrieval as used in \cite{Seidel2020}. Further details on degeneracies, line broadening types, and the continuum parameter can be found in the first paper on MERC \cite{Seidel2020}, and in the works the forward model is based on \citep{Ehrenreich2006, Pino2018}.

\section{Retrieval results}
\label{sec:results}

We applied MERC with its updates presented in Section \ref{sec:MERC} to the combined HARPS and ESPRESSO dataset presented in Section \ref{sec:dataset}. We examined multiple forward models, all with added planetary rotation: an isothermal temperature profile (iso), a temperature gradient (T grad), both with no added winds, and an isothermal temperature profile combined with the following wind patterns: 
\begin{itemize}
\item uniform day-to-night side wind (dtn)
\item $\cos\theta$ dependent day-to-night side wind ($\mathrm{dtn}_{\cos\theta}$)
\item uniform super-rotational wind (srot)
\item $\cos\theta$ dependent super-rotational wind ($\mathrm{srot}_{\cos\theta}$)
\item vertical wind (ver)
\item two layer approach with either a super-rotational or day-to-night side wind in the lower atmosphere combined with a vertical wind in the upper atmosphere. The switch between atmospheric layers is tested at different pressures. ($\mathrm{dtn}_{\cos\theta}, \mathrm{ver}$; dtn, ver; srot, ver)
\end{itemize}

One of the most prominent features of Bayesian retrieval is the application of priors. Priors set the boundary conditions for the parameters space and allow to add knowledge of the physical properties to the numerical analysis of the fits. The priors used in this work are listed in Table \ref{table:priors}. The priors for the mean isothermal temperature T$_{\mathrm{iso}}$ and the degenerate continuum parameter NaX were set to exclude non-physical results in terms of temperature, pressure and sodium abundance. The velocity parameters were set more freely, from the lowest value close to zero to velocity ranges far beyond the escape velocity. The velocity priors are set each in both directions and have no sign. The impact of the priors will be discussed in Section \ref{sec:results}. The plotting range of the posteriors in Appendix \ref{app:nolat} is selected for visibility and does not reflect the entire prior range.

\begin{table*}
\caption{Overview of the different models' prior ranges.}
\label{table:priors}
\centering
\begin{tabular}{l c c c c c}
\hline
\hline
Model & T$_{\mathrm{iso}}$ [K]  &  NaX & v$_{\mathrm{srot}}$ [\kms] & v$_{\mathrm{dtn}}$ [\kms] & v$_{\mathrm{ver}}$ [\kms]     \\
\hline
isothermal  & [1500, 4000]    & [-5.0, -1.0]  & -  & -  &  - \\
T gradient  & [800, 4000],[2000, 6000]\tablefootmark{a}    & [-5.0, -1.0]  &  - & -  & - \\ 
dtn  &  [1500, 4000]   & [-5.0, -1.0]  & -  & [0.1,40.0]  & -\\
$\mathrm{dtn}_{\cos\theta}$  &  [1500, 4000]   & [-5.0, -1.0]  & -  &  [0.1,40.0]  & -\\
srot  &  [1500, 4000]   & [-5.0, -1.0]  & [0.1,40.0]   &  - & -\\
$\mathrm{srot}_{\cos\theta}$  &  [1500, 4000]   & [-5.0, -1.0]  & [0.1,40.0]   &  - & - \\
ver  &   [1500, 4000]  & [-5.0, -1.0]  & -  &  - & [0.1,40.0] \\
srot, ver  &  [1500, 4000]   & [-5.0, -1.0]  &  [0.1,20.0] & -  & [0.1,40.0]\\ 
dtn, ver  &   [1500, 4000]  & [-5.0, -1.0]  & -  & [0.1,20.0]  & [0.1,40.0]\\
$\mathrm{dtn}_{\cos\theta}, \mathrm{ver}$ &  [1500, 4000]   & [-5.0, -1.0]  &  - & [0.1,20.0]  & [0.1,40.0] \\
\hline
\end{tabular}
\tablefoot{\tablefoottext{a}{Bottom and top temperature respectively to create a temperature gradient.}}
\end{table*}

\noindent The isothermal model with no added winds (but taking into account planetary rotation) is the simplest model and serves as our base to calculate the significance of the Bayesian evidence for all other models with $|\ln\mathcal{Z}|=2111.55\pm0.29$. The Bayesian evidence of each model, as well as the significance of the model according to the Jeffrey's scale in Table \ref{table:Bayesianoverview} are shown in Table \ref{table:comparison} with a visual representation of the strength of evidence in Figure \ref{fig:visual}. The posteriors from the nested sampling retrieval as corner plots are in Appendix \ref{app:nolat}, with the mean values on the top of each column and the parameters retrieved with the highest Bayesian evidence marked in blue.

\begin{table}
\caption{Comparison of the different models. The base model to calculate $|\ln\mathcal{B}_{01}|$ is the isothermal model with no added wind patterns. The comparison stems from the Jeffrey's scale in Table \ref{table:Bayesianoverview}. The model with the highest Bayesian evidence is highlighted in bold.}
\label{table:comparison}
\centering
\begin{tabular}{l c c l}
\hline
\hline
Model & $|\ln\mathcal{Z}|$   &  $|\ln\mathcal{B}_{01}|$ & Strength of evidence    \\
\hline
isothermal  &  $2111.55\pm0.29$   & -  & - \\
T gradient  &  $2110.72\pm0.30$   & $-0.83$  & No evidence \\ 
dtn  &  $2114.23\pm0.39$   & $2.68$  & Moderate evidence \\
$\mathrm{dtn}_{\cos\theta}$  &  $2112.46\pm0.35$   & $0.91$  & No evidence \\
srot  &  $2116.00\pm0.38$   & $4.45$  & Moderate evidence \\
$\mathrm{srot}_{\cos\theta}$  &  $2115.87\pm0.37$   & $4.32$  & Moderate evidence \\
ver  &  $2124.07\pm0.34$   & $12.52$  & Strong evidence \\
srot, ver  &  $2122.86\pm0.36$   & $11.31$  & Strong evidence \\ 
\textbf{dtn, ver}  &  $2125.17\pm0.39$   & $13.62$  & Strong evidence \\
$\mathrm{dtn}_{\cos\theta}, \mathrm{ver}$  &  $2123.82\pm0.32$   & $12.27$  & Strong evidence \\
\hline
\end{tabular}
\end{table}

\begin{figure}[htb!]
\resizebox{\columnwidth}{!}{\includegraphics[trim=3.0cm 3.0cm 5.0cm 2.5cm]{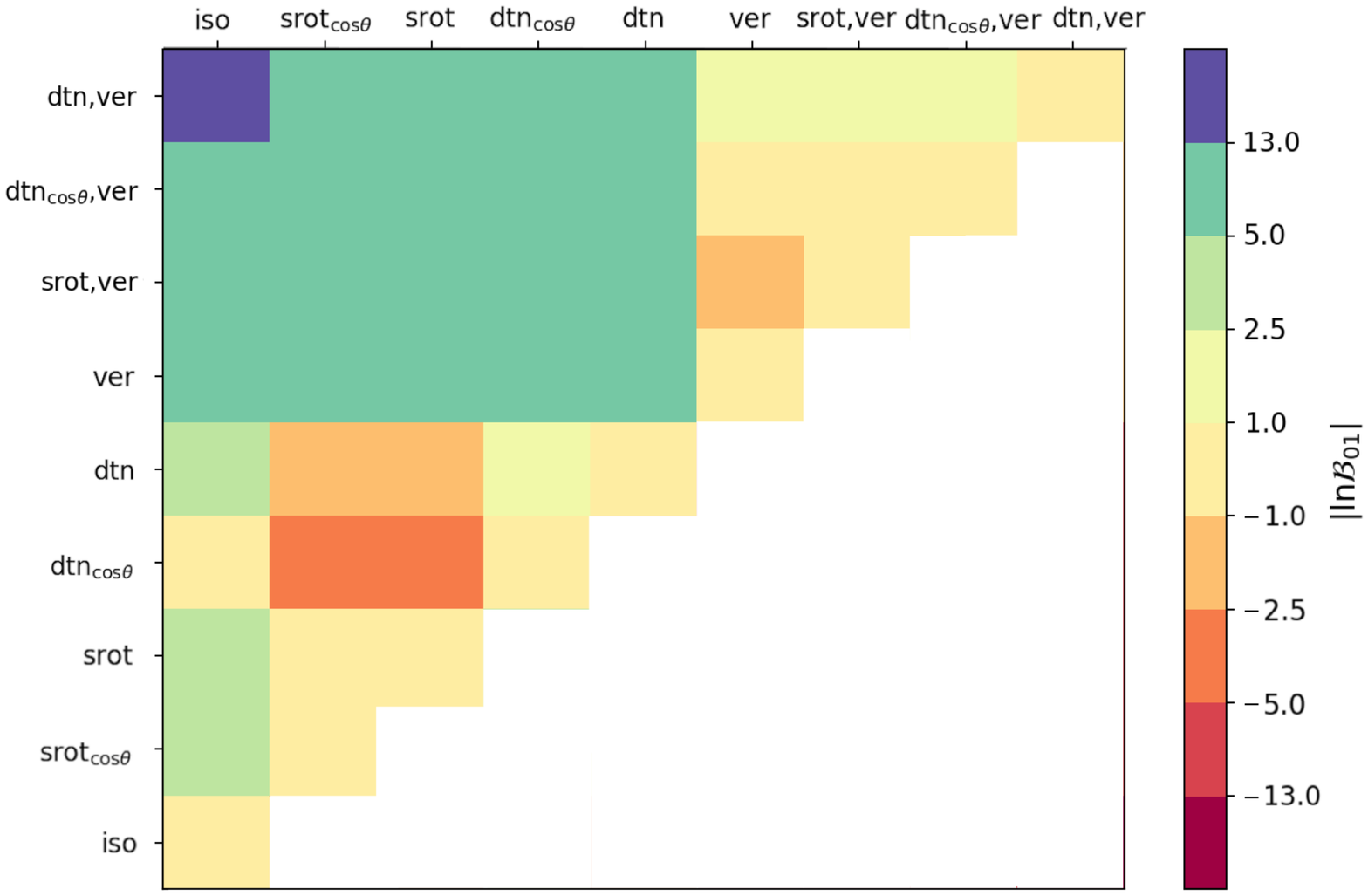}}
        \caption{Visualisation of the difference in Bayesian evidence of Table \ref{table:comparison}. The colour scale indicates $|\ln\mathcal{B}_{01}|$ with the colours set at the different limits of the Jeffrey's scale (Table \ref{table:Bayesianoverview}), with strong evidence in green and the overall highest evidence in purple. The temperature gradient model is left out of the visualisation, because it has a lower Bayesian evidence than even the base model. The model with the highest evidence (uniform day-to-night side wind in lower and vertical wind in the upper atmosphere) shows strong evidence compared with all rotational models and moderate evidence with the vertical model and all two-layer models, see top row. For the exact values, see Table \ref{table:comparison}.}
        \label{fig:visual}
\end{figure}

\subsection{Temperature profiles}

Comparing the Bayesian evidence of the temperature gradient model to the isothermal model shows that there is not sufficient evidence for one model over the other. The temperature gradient was approximated via a base and a top temperature, where the base is the continuum level of the data and the top the highest probed layer of the atmosphere. In fact, the temperature gradient model converges towards the isothermal solution and is then punished for the additional parameter with lower Bayesian evidence (see Figure \ref{fig:gradientposterior}). We have, therefore, used an isothermal temperature profile throughout all other retrieved models, which should only be interpreted as an upper limit mean temperature in light of the variation of the datasets in line depths and the intrinsic nature of transit observations as a mean over the terminator. The application of a homogeneous mean temperature profile does, however, not rule out the existence of a temperature gradient, which might be temporally variable and averaged over in the combined dataset. The combined dataset is needed to properly constrain winds in the atmosphere, the main goal of this work, and a more in-depth treatment of temporal temperature variations in the atmosphere of WASP-76~b is deferred to future work.
The range of retrieved temperatures for all models is higher than the equilibrium temperature of WASP-76~b, the approximate temperature of $3300-3400$ K for the hottest part in the higher layers of the atmosphere retrieved for the here presented models is even higher than expected from \citep{Ehrenreich2020}. The temperature retrieval is driven by the line depth and thus influenced by the level of the continuum and the accuracy of the line depth and system parameters. In MERC, we take into account Rayleigh scattering but do not account for the effects of $H^-$ opacity \citep{Gandhi2020} on the continuum level for computational reasons. Therefore, the temperature should be interpreted as an upper boundary only. However, even as an upper limit, the temperature remains curiously high and the impact and likely cause are discussed in Section \ref{sec:diss}. 

\subsection{Latitude dependent wind patterns}

Of all models presented in this section, only those with a day-to-night side wind or a super-rotational wind could show a latitude dependence. In most studies using GCMs these wind patterns are restricted to bands along the equator (called jets) and weaken towards the poles  \citep{Showman2009,Showman2018,Parmentier2018}. We approximate this behaviour by introducing a solid-body rotation dependence ($\cos\theta$) to the wind pattern as described in Section \ref{sec:wind}. The wind speeds retrieved for all models in this section ($2-4\, \kms$) are  compatible with GCM studies \citep[e.g.][]{Showman2018, Parmentier2018}. Comparing the Bayesian evidence between the uniform wind patterns and their $\cos\theta$ counterparts shows no evidence for a better fit with a jet-like structure for super-rotational wind ($|\ln\mathcal{B}_{01}|=0.13$), weak evidence for a better fit with a uniform wind pattern for the day-to-night side wind over a jet-like structure  ($|\ln\mathcal{B}_{01}|=1.77$), and weak evidence for a better fit when the uniform wind pattern is applied to the lower atmosphere ($|\ln\mathcal{B}_{01}|=1.36$) in the combined wind patterns (lower day-to-night-side wind and upper vertical wind). Although the evidence for a preference of the uniform wind patterns with no latitude dependence is only weak (see Table \ref{table:Bayesianoverview}), it shows that for WASP-76~b the winds are likely not primarily restricted in jets and flow more evenly across the terminator. However, this does not rule out any jet-like structure in the lower atmosphere of WASP-76~b, especially a multi-jet scenario as seen for Jupiter \citep{LiuSchneider2010} that would look identical to a wide jet when integrated over the atmosphere.

\subsection{Uniform wind patterns}

All tested uniform wind patterns show moderate to strong evidence for a better fit when compared to the base model with no winds (see Table \ref{table:comparison}). A vertical wind throughout the atmosphere at $30.8^{+3.1}_{-3.1}\, \kms$ is not only preferred over the base model ($|\ln\mathcal{B}_{01}|=12.52$, strong evidence), but also over the next best wind pattern ($|\ln\mathcal{B}_{01}|=8.07$, strong evidence). This indicates that a strong vertical wind is needed to create the broadened line shape which cannot be created solely from a moderate day-to-night side or super-rotational wind at the retrieved velocities of $2-4\, \kms$.

Comparing the day-to-night side wind with the super-rotational wind, it appears at first glance that the super-rotational wind is preferred ($|\ln\mathcal{B}_{01}|=1.77$, weak evidence). This serves as an excellent reminder as to not being aware of model limitations and blind application of nested sampling. The super-rotational broadening splits the sodium doublet into two sub-peaks, one red-shifted at the morning terminator, one blue-shifted at the evening terminator, which creates a strong broadening in the wings. Consequently, every part of the atmosphere receives some shift and the line depth is severely decreased by a strong super-rotational wind. While this would be an excellent fit for a very broad but shallow sodium doublet, it does not work similarly for the deep sodium lines in this dataset. To subsequently offset the lack of un-shifted parts of the atmosphere which would sit at the centre, the retrieval increases the temperature, which is the main driver of the line depth. This leads to retrieved temperatures well beyond $4000$ K, which are not physically sensible for an ultra hot Jupiter with the system parameters of WASP-76~b (see Table \ref{tab:para} and posterior distributions in Appendix \ref{app:nolat}) and are ruled out via the prior. Additional evidence for this line of thought comes from the combined two layer models, where part of the atmosphere has the sodium lines un-shifted, but broadened, by vertical wind, and part of the atmosphere is dominated by either a day-to-night side or a super-rotational wind. In this scenario, the day-to-night side wind is preferred over the super-rotational wind, as predicted ($|\ln\mathcal{B}_{01}|=2.31$, moderate evidence within the uncertainty). With ESPRESSO, the resolution would be sufficient to show this split into two peaks of the sodium line, provided that the SNR is sufficient and that super-rotation is the dominant wind pattern in the atmosphere and not super-imposted with other wind patterns that leave parts of the atmosphere at net zero wind speed (e.g. a combination with day-to-night side winds).

\subsubsection{Two layer patterns}

\begin{figure*}[htb!]
\resizebox{\textwidth}{!}{\includegraphics[trim=1.0cm 0.0cm 1.0cm 0.0cm]{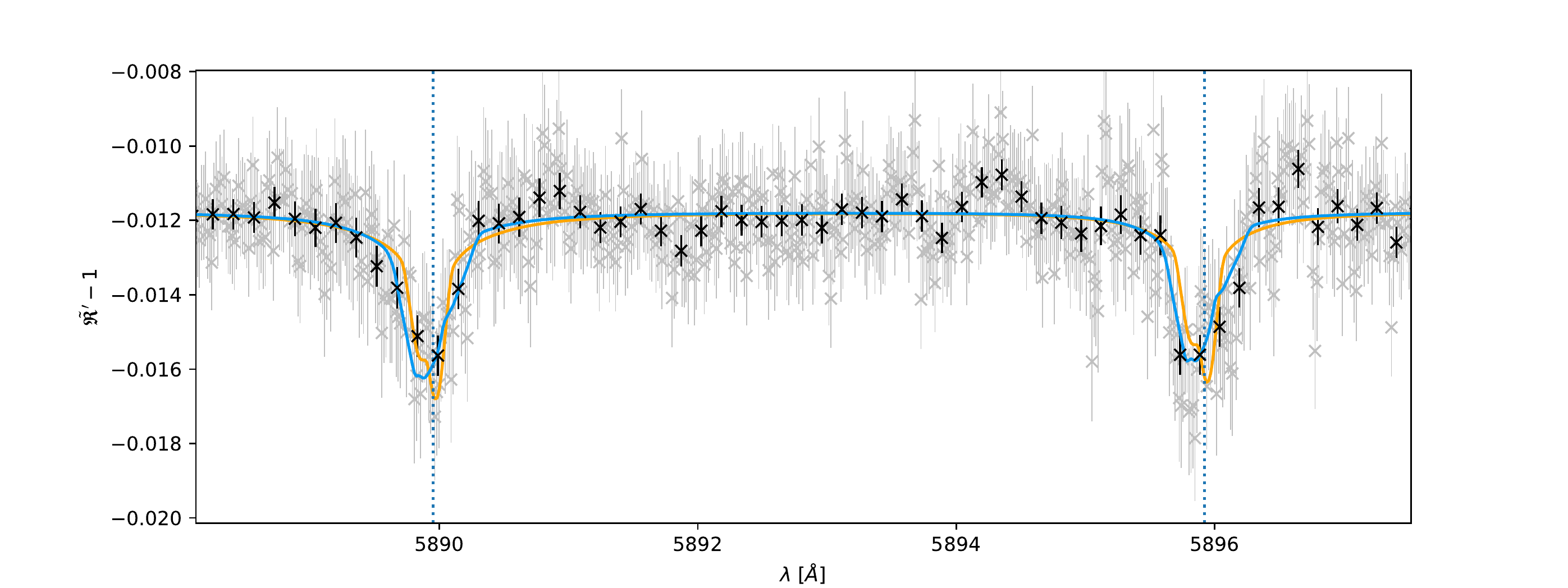}}
        \caption{Best-fit of isothermal line retrieval with an added day-to-night side wind in the lower atmosphere with no $\cos\theta$ dependence and a vertical wind in the upper atmosphere in blue on the dataset in gray. The day-to-night side wind with no $\cos\theta$ dependence throughout the atmosphere (moderate evidence) is shown for comparison purposes in orange. According to the Jeffrey's scale, which is logarithmic, the best-fit model in blue has more than an order of magnitude higher probability than the moderate evidence model in orange. The data was binned by 10 bins in black for better visualisation, the sodium doublet line centre are indicated by dashed, vertical, blue lines.}
        \label{fig:dtn_verbestfit}
\end{figure*}

To approximate a more realistic structure of exoplanet atmospheres, we divide the atmosphere into a lower and higher atmosphere and allow for different wind patterns in the two layers. In the lower layer, we apply a day-to-night side or a super-rotational wind and a vertical wind in the higher layer. The first question is where to set the boundary between the two layers. As shown in \cite{Seidel2020}, the additional parameter for the switch of layers introduces too many variables not allowing for proper convergence of the retrieval. Luckily, given the log-scale of the pressure, which serves as a proxy for relative height above the surface, only a few values for the layer switch are sensible and a restriction beyond an order of magnitude estimation isn't necessary, especially since there will most likely be a transition zone between the wind patterns. Assuming a surface pressure at the continuum of $10$ bar as is customary in the literature \citep[e.g.][]{LecavelierdesEtangs2008, Agundez2014,Line2014, Pino2018}, we explore layer switches at $10^{-3}$ bar, $10^{-4}$ bar and $10^{-5}$ bar. $10^{-3}$ bar compared to the continuum pressure, as the lowest value, corresponds to the location of the jets found for hot Jupiters via GCM modelling \citep{Showman2009}, therefore, we know a priori that zonal winds dominate the atmosphere below this threshold. As is evident from the posterior plots for a layer switch at $10^{-5}$ bar (see Figure \ref{fig:dtn_verbestposterior4} and \ref{fig:srot_ver4}), this switch is set too high and cannot properly constrain the upper layer. The preferred switch of layers, if any, should therefore lie between these two values. The presence of a switch between dynamical regimes at $10^{-3}$ bar, as predicted by GCM results, is preferred with at least $|\ln\mathcal{B}_{01}|=4.56$ (strong evidence within the uncertainty of the results) independently of the assumed dynamics in the lower layer (dtn, srot, or ver). We thus fix the pressure switch at $10^{-3}$ bar for the two layer pattern models. 

\begin{figure}[htb!]
\resizebox{\columnwidth}{!}{\includegraphics[trim=1.0cm 0.0cm 1.0cm 0.0cm]{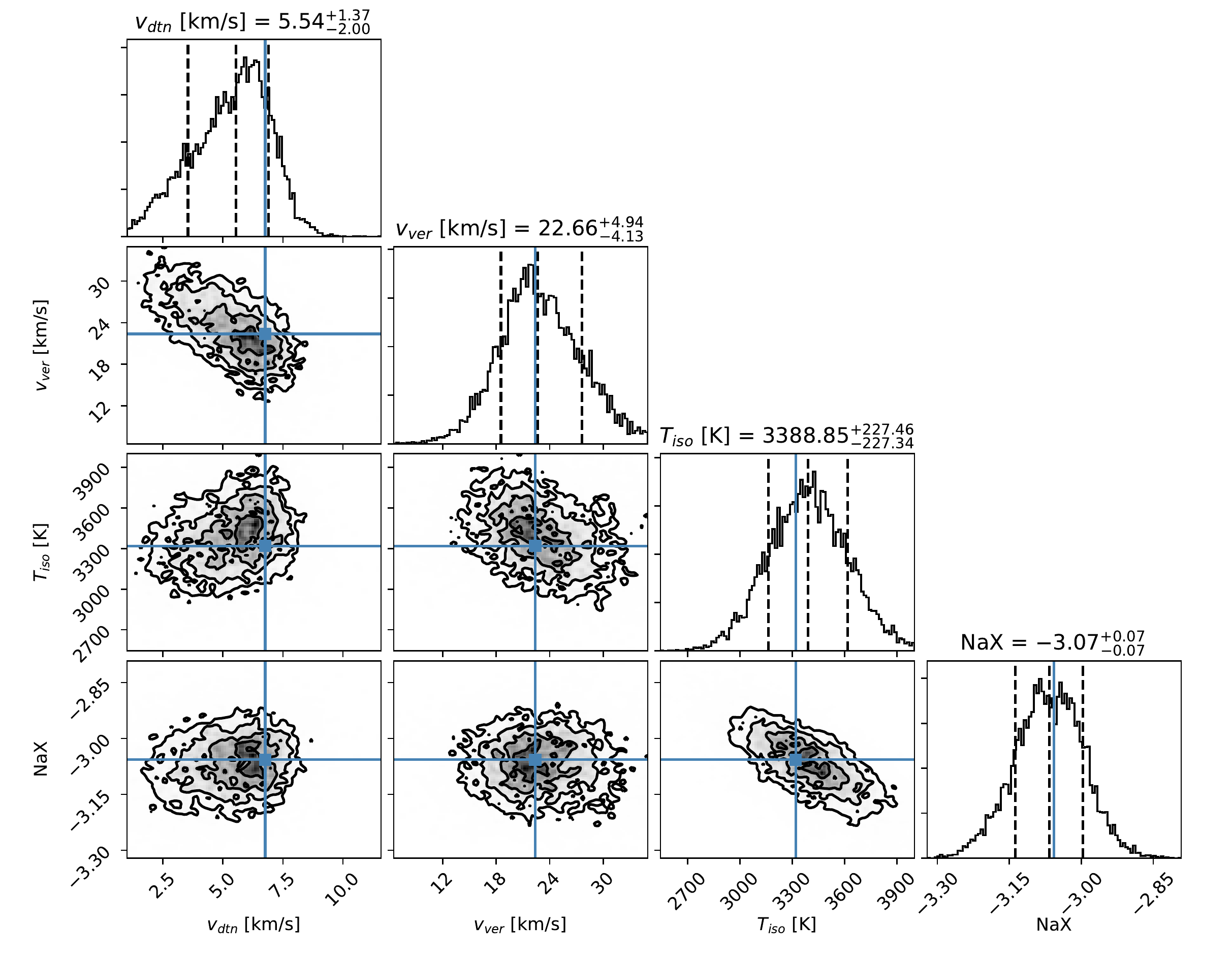}}
        \caption{Posterior distribution of isothermal line retrieval with an added day-to-night side wind in the lower atmosphere and a vertical wind in the upper atmosphere. The change of layers was set at $p=10^{-3}$ bar, with the pressure at the surface set to $P_0=10$ bar.}
        \label{fig:dtn_verbestposterior}
\end{figure}

Of these models, a uniform day-to-night side wind has the highest evidence in the lower atmosphere, with moderate evidence over the super-rotational wind and weak evidence over a $\cos\theta$ latitude dependence of the wind, which are the next best two models. Therefore, the model with vertical wind in the upper and day-to-night side wind in the lower atmosphere is prefered over any other model. The preference for a uniform wind in the lower atmosphere over the $\cos\theta$ latitude dependence implies that if the winds in the lower atmosphere in WASP-76~b are restricted to a jet-like band at the equator, it has to be relatively broad, spanning at least farther in latitude than a scaling via solid-body rotation suggests ($>30^\circ$). The higher evidence for a day-to-night side wind over a super-rotational wind stems from a visible offset of the sodium doublet to the blue, as highlighted in \cite{Tabernero2020} where a median offset of $-5.4\pm 2.6$ for all detected lines was found. The posterior distribution of the best-fit model is shown in Figure \ref{fig:dtn_verbestposterior} and the resulting retrieved transmission spectrum overplotted on the data in Figure \ref{fig:dtn_verbestfit}. The day-to-night side wind has a speed of $5.5^{+1.4}_{-2.0}\, \kms$, while the vertical wind shows speeds of $22.7^{+4.9}_{-4.1}\, \kms$. Since we have taken into account planetary rotation under the assumption of tidal locking, these wind speeds are not upper limits, but constrain the possible wind speeds in the atmosphere of WASP-76~b to within $30\%$ for the lower layer and $20\%$ for the upper layer (see uncertainty of the retrieval). However, as discussed in Section \ref{sec:combined}, due to the necessary exposure time, the lines will be smeared by one resolution element in each direction, which increases the error on the wind speed for the radial outwards wind by roughly $10\%$ (dominated by the width) and roughly $10\%$ as well for the day-to-night side wind (lower absolute wind speed, but dominated by the line position). This has no influence on the overall conclusions on the wind patterns present in the studied system.

\section{Discussion}
\label{sec:diss}

\noindent Recently, \cite{Ehrenreich2020} studied WASP-76~b using ESPRESSO via the atomic iron signature and found significant asymmetry between the first and second half of the transit, leading to a one-sided shift of the signature. At first, the iron detection shows no shift during ingress and then moves to a constant blue-shift of $11.0\pm0.7\,\kms$. Taking into account planetary rotation, they conclude that the lower atmosphere of WASP-76~b shows a day-to-night side wind of $5.3\, \kms$. The result, a best-fit to the sodium doublet with a day-to-night side wind of $5.5^{+1.4}_{-2.0}\, \kms$, is compatible with their findings in the lower atmosphere where iron is present (at pressures higher than the micro bar level). This result is driven by the detected offset of the lines to the blue by $\sim 5\, \kms$ in \cite{Tabernero2020}, which is confirmed in this work and can either be induced by a day-to-night side wind over the entire planet or by a super-rotational wind with condensation of sodium on the night side. Given the condensation temperature of sodium ($50\%$ condensated with T $\approx 1000$~K; \citealt{Lodders2003}) is much too low for the night side temperature of an ultra hot Jupiter \citep{Parmentier2018}, we rule out an asymmetrical super-rotational wind. A similar offset observation was made for WASP-121~b, another ultra hot Jupiter \citep{Borsa2020}. In the here presented analysis, we neglected the impact of the RM effect, which was taken into account in \citet{Ehrenreich2020}. The verification of the wind pattern from \citet{Ehrenreich2020} thus additionally confirms the negligible impact of the RM effect for this particular system.

None of the studies analysing wind patterns in the lower atmosphere are, however, capable to probe what happens in the intermediate atmosphere, between the mass-loss in the exosphere and the various zonal wind patterns in the lower atmosphere. We show via atmospheric retrieval that these layers are connected via a strong vertical wind of $22.7^{+4.9}_{-4.1}\, \kms$ for WASP-76~b. In \cite{Seidel2020}, a similar phenomenon was observed for HD189733~b and later confirmed via independent observations in \cite{Keles2020}. This vertical wind can either be an outwards expanding wind, where the material does not descend back towards lower atmospheric layers or the vertical parts of large cell-like structures, where the upper, zonal arm of the cell is in the thermosphere. Assuming this large cell-like structure, the upper, zonal arm would not influence the sodium line shape, because sodium would be ionised at these altitudes and only recombine in the downwards vertial arm of the cell. Due to the nature of transmission observations over the terminator, these two wind patterns cannot be distinguished. In the case of HD189733~b, the interaction of ions in the lower atmosphere with the surprisingly strong estimated magnetic field \citep{Cauley2019} could be a possible explanation for a purely expanding wind \citep{Seidel2020}, dragging neutral sodium upwards. For WASP-76~b, no indirect magnetic field strength estimates are available, however, if the same mechanism is at work for this exoplanet, the magnetic field strength should be comparable to the magnetic field strengths observed in the sample of \cite{Cauley2019}, orders of magnitude higher than the magnetic field of Jupiter.

The observed vertical wind, irrespective of its origin, feeds material up to the thermosphere, where it could be further heated and escape via hydrodynamic escape. This has not been directly observed for WASP-76~b, but strong planetary winds, bordering on hydrodynamical escape are predicted for WASP-76~b (see \cite{Salz2016b}, Figure 3, for $\log_{10}\Phi_G (\mathrm{WASP-76~b}) \approx 13$). This is compatible with the results in \cite{Seidel2019}, that showed sodium expands far above the surface, but velocities remain just below the escape velocity for WASP-76~b. The revised escape velocity with the updated planetary radius and mass is $41\pm1 \kms$, significantly higher than the vertical wind speeds retrieved for sodium. Therefore, the vertical wind is a transport mechanism for material to the thermosphere but not the direct driver of atmospheric escape. However, it is important to note that due to the integration over the entire atmosphere, variations of the wind speed within the two atmospheric regimes are not accounted for, but rather a mean wind velocity is derived. Whether the expanding wind gains momentum with height, driven by processes in the thermosphere, or fuels it themselves is as a consequence unclear.

While the result provides compelling evidence for the case of strong vertical winds in the upper atmosphere, there are competing ideas to explain the broad line shape of sodium in hot Jupiters: For low sodium abundance, the sodium might form an optically thin cloud high up in the atmosphere, producing a very broad feature. However, this is dependent on the line ratio and has been shown to not apply to WASP-76~b \citep{OzaGebek2020}. Additionally, more complex temperature profiles might apply for WASP-76~b, varying in latitude or longitude, which the retrieval cannot account for. While the line shape is sensitive to the temperature via thermal broadening, it cannot produce the line shape seen for the sodium doublet in hot Jupiters \citep{Wyttenbach2015, Huang2017, Seidel2020}. Another caveat is the assumption of local thermodynamic equilibrium (LTE) which excludes any non-LTE effects that might influence the line shape. Recently, in \cite{Fisher2019} this assumption was quantified, highlighting that as of now, LTE and non-LTE atmospheres cannot be distinguished, which strengthens the case for strong vertical winds. \\

 \cite{Zhang2020} recently reviewed current knowledge regarding temperature profiles for ultra hot Jupiters (defined as having equilibrium temperatures higher than $\sim 2200$~K). Temperature inversions were confirmed via emission observations for KELT-9~b \citep{Pino2020}, WASP-121~b \citep{Evans2017, Bourrier2020}, a planet similar to WASP-76~b, for WASP-18~b \citep{Sheppard2017}, for WASP-189~b \citep{Yan2020}, and WASP-33~b \citep{Haynes2015,Nugroho2017}. Nonetheless, temperature inversions are not established as an unequivocal feature of ultra hot Jupiters, given that some, for example, WASP-12~b and WASP-103~b, have spectra consistent with blackbody radiation, suggesting isothermal atmospheres \citep{Arcangeli2018, Parmentier2018, Kreidberg2018}. As a caveat, these observations stem from WFC3 and the blackbody spectra in the covered band could also be explained with a temperature inversion, which might attribute these outliers to observational limitations. TiO and VO have been suggested as two possible drivers of thermal inversion by heating the upper atmosphere \citep{Hubeny2003, Fortney2008}, although discussions on the mechanisms of thermal inversions are still ongoing \citep{Molliere2015, Lothringer2019, Gandhi2019} and atomic metals were suggested as an alternative inversion agent (e.g. ZrO) \citep{Lothringer2019,Tabernero2020}. This additional caveat is especially important given that some of the mentioned works on temperature inversions are disputed, for example in WASP-33~b, where the features seen in \cite{Nugroho2017} could not be confirmed by \cite{Herman2020}, but were again seen tentatively in \cite{Serindag2020}. \cite{Nugroho2020} even found iron lines in emission, strengthening the case for additional causes of inversion layers apart from TiO and VO. Assuming TiO and VO as the main drivers of inversion layers, in \cite{Tabernero2020},  the presence of TiO and VO was ruled out down to a level of $<10$ ppm and, in line with the reasoning from \cite{Nugroho2017}, they suggest that TiO and VO might be hidden under stronger features or very weak, with another possibility being that both are transported from the day-side to the night-side by a day-to-night-side wind where they condense out of the atmosphere. While the retrieval presented here cannot, and does not aim to, shed light on the detailed temperature structure of WASP-76~b, an isothermal temperature profile is not at odds with current knowledge on ultra hot Jupiters and the suggestion of day-to-night side winds from \cite{Tabernero2020} to explain their findings on TiO and VO is consistent with our results on winds.
 
As already mentioned in Section \ref{sec:results}, the temperature retrieved here is surprisingly high when compared to the equilibrium temperature and would lead to the ionisation of sodium, which we estimated with FastChem \citep{Stock2018}. At the equilibrium temperature of the planet, sodium remains in its neutral state at fairly constant abundance up to $10^{-5}$~bar, whereas at temperatures of $3300$~K and beyond sodium ionises far lower in the atmosphere, in disagreement with the detected and confirmed sodium feature. This apparent inconsistency, where the retrieved temperature implicates an improbably high initial sodium abundance to compensate for ionisation, merits a closer look.  Instead of concluding that an unphysical initial sodium abundance has to be present in ultra hot Jupiters, it is more likely that vertical winds, such as found here, that interact with the sodium in the atmosphere inevitably alter the density profile of sodium, away from a hydrostatic description of the atmosphere. In practise, the modified density profile will inflate the line depth which has to then be compensated for with a higher homogeneous sodium abundance in the hydrostatic model. In conclusion, while a change in density profile has no significant impact on the line shape and, therefore, the here presented results, winds will modify density profiles and studies hoping to fit sodium abundances, in low or high resolution, has to account for the density profile of sodium generated by mid-latitude winds such as detected in this work.

\section{Conclusions}
\label{sec:conl}
 
We have studied winds in the atmosphere of the ultra hot Jupiter WASP-76~b via its resolved sodium doublet. The transmission spectrum in the wavelength range of the sodium doublet is composed of two nights of ESPRESSO data \citep{Ehrenreich2020,Tabernero2020} and the reprocessed HARPS dataset published in \cite{Seidel2019}, which were combined via S/N weighing. We introduce an update to the MERC code from \cite{Seidel2020}, which now takes planetary rotation into account and allows latitude dependent zonal wind patterns. We found no evidence that a temperature gradient is preferred over an isothermal atmosphere. We combined the isothermal temperature profile with a uniform day-to-night side wind, a latitude dependent day-to-night side wind, a uniform super-rotational wind, a latitude dependent super-rotational wind, a vertical wind and a two layer combination of either day to night side or super-rotational winds in the lower and the vertical wind in the upper atmosphere. 

The best-fit model is a uniform day-to-night side wind in the lower atmosphere of $5.5^{+1.4}_{-2.0}\, \kms$ with a vertical, most likely mainly expanding wind in the upper atmosphere of $22.6^{+4.9}_{-4.1}\, \kms$. Assuming a surface pressure of $P_0 = 10$ bar, the switch between the two layers is set at $p=10^{-3}$ bar. For all models, we retrieve an isothermal temperature between $3300-3400$ K and discuss that this overestimate is most likely a direct result of the impact of winds on the density profile of sodium in WASP-76~b, an important caveat for future studies regarding the fit of abundances from spectral lines. Our findings are compatible with the current literature on the wind dynamics of lower atmospheres and temperature profiles of hot and ultra hot Jupiters, especially with previous work on WASP-76~b. \cite{Ehrenreich2020},  using the time-resolved iron detection with ESPRESSO, found a day-to-night side wind of $5.3\, \kms$, a result corroborated by our findings in this study. Additionally, we demonstrate a need for vertical winds in the intermediate atmosphere of these highly irradiated gas giants via their broadened features.

Our work proves the effectiveness of direct retrieval on resolved spectral lines to constrain possible wind patterns as input for more in-depth theoretical studies and the crucial role of high-resolution spectrographs for atmospheric characterisation. Our findings show a clear need for more detailed climate simulations taking into account the higher layers of the atmosphere to draw a clear picture of the atmospheres of ultra hot exoplanets.

\begin{acknowledgements}
This project has received funding from the European Research Council (ERC) under the European Union's Horizon 2020 research and innovation programme (project {\sc Four Aces}; grant agreement No. 724427).
This work has been carried out within the frame of the National Centre for Competence in Research `PlanetS' supported by the Swiss National Science Foundation (SNSF). This work was supported by FCT - Funda\c{c}\~ao para a Ci\^encia e Tecnologia (FCT) through national funds and by FEDER through COMPETE2020 - Programa Operacional Competitividade e Internacionaliza\c{c}\~ao by these grants: UID/FIS/04434/2019; UIDB/04434/2020; UIDP/04434/2020; PTDC/FIS-AST/32113/2017 \& POCI-01-0145-FEDER-032113; PTDC/FIS-AST/28953/2017 \& POCI-01-0145-FEDER-028953. VA acknowledges the support from FCT through Investigador FCT contract nr.  IF/00650/2015/CP1273/CT0001. NJN acknowledges support form the following projects: UIDB/04434/2020 \& UIDP/04434/2020, CERN/FIS-PAR/0037/2019, PTDC/FIS-OUT/29048/2017, COMPETE2020: POCI-01-0145-FEDER-028987 \& FCT: PTDC/FIS-AST/28987/2017. The INAF authors acknowledge financial support of the Italian Ministry of Education, University, and Research with PRIN 201278X4FL and the "Progetti Premiali" funding scheme. JLB acknowledges financial support from ”la Caixa” Foundation (ID  100010434) under the fellowship LCF/BQ/PI20/11760023, and from the European Union's Horizon 2020 research and innovation programme under the Marie Sk\l odowska-Curie grant agreement No 847648. AW acknowledges the financial support of the SNSF by grant number P400P2$\_$186765. SGS acknowledges the support from FCT through Investigador FCT contract nr. CEECIND/00826/2018 and POPH/FSE (EC). ODSD is supported in the form of a work contract (DL 57/2016/CP1364/CT0004) funded by FCT. 
We would like to thank the anonymous referee for their thorough and thoughtful comments that greatly improved the quality of the manuscript.
\end{acknowledgements}

%
\bibliographystyle{aa} 
\bibliography{IntoTheStorm}
%
\appendix
\section{Posterior distributions of the retrievals}
\label{app:nolat}
\begin{figure}[hbt]
\resizebox{\columnwidth}{!}{\includegraphics[trim=-0.0cm 0.0cm -0.0cm 0.0cm]{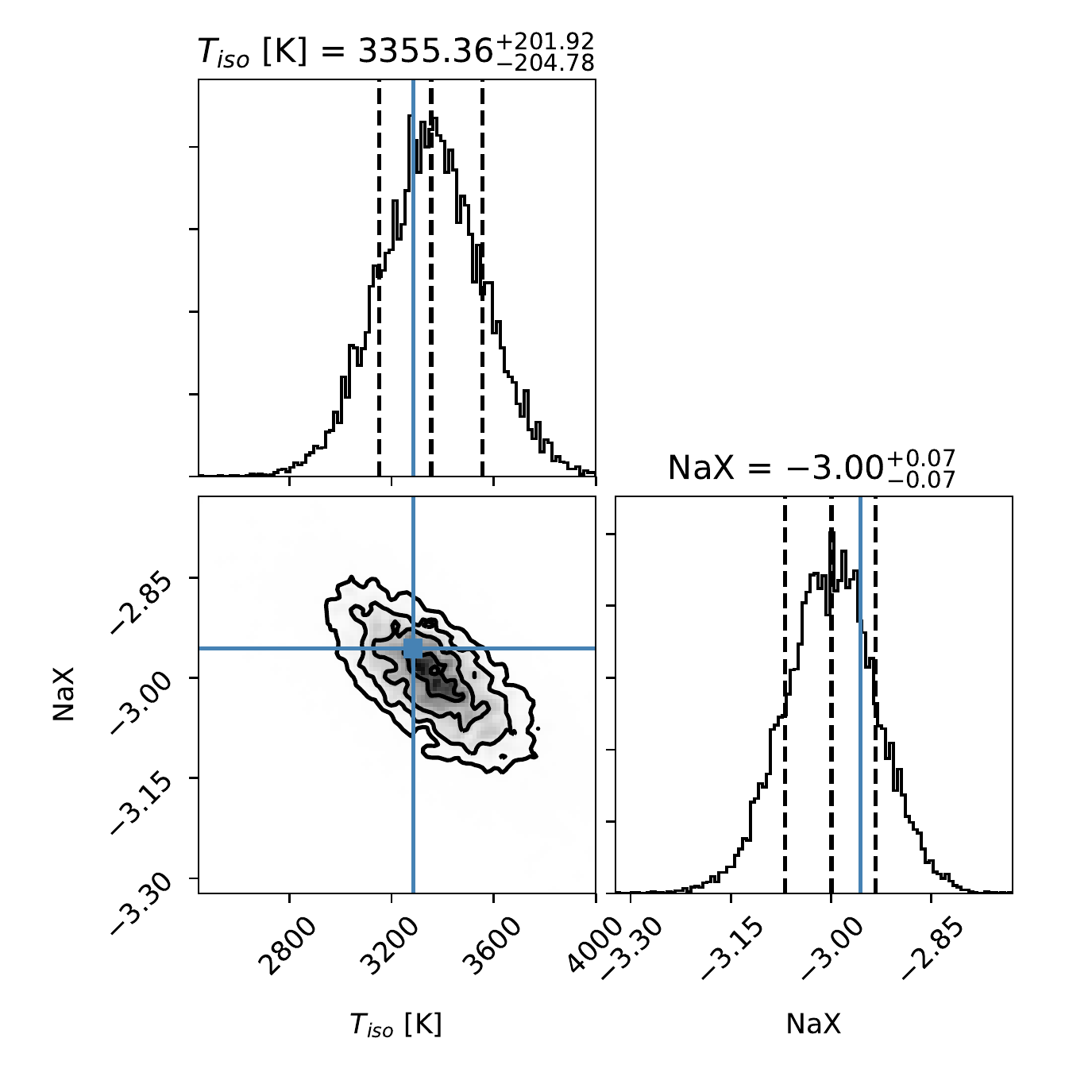}}
        \caption{Posterior distribution of isothermal line retrieval. This was used as the base model to compare all other scenarios to.}
        \label{fig:isoposterior}
\end{figure}

\begin{figure}[htb!]
\resizebox{\columnwidth}{!}{\includegraphics[trim=0.0cm 0.0cm 0.0cm 0.0cm]{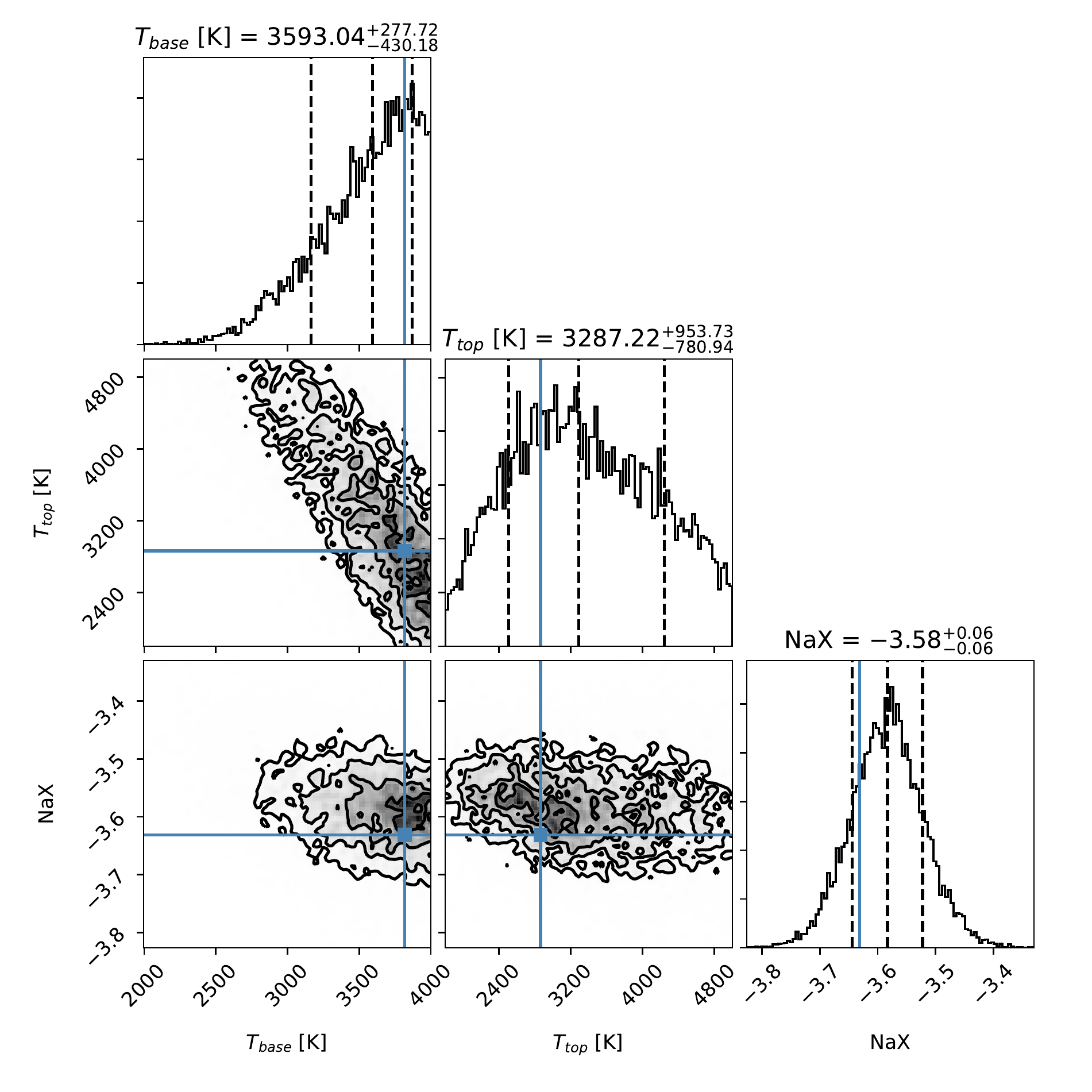}}
        \caption{Posterior distribution of temperature gradient line retrieval. $T_{\mathrm{base}}$ is at the surface of the planet and $T_{\mathrm{top}}$ at the top of the probed atmosphere. }
        \label{fig:gradientposterior}
\end{figure}

\begin{figure}[htb!]
\resizebox{\columnwidth}{!}{\includegraphics[trim=0.0cm 0.0cm 0.0cm 0.0cm]{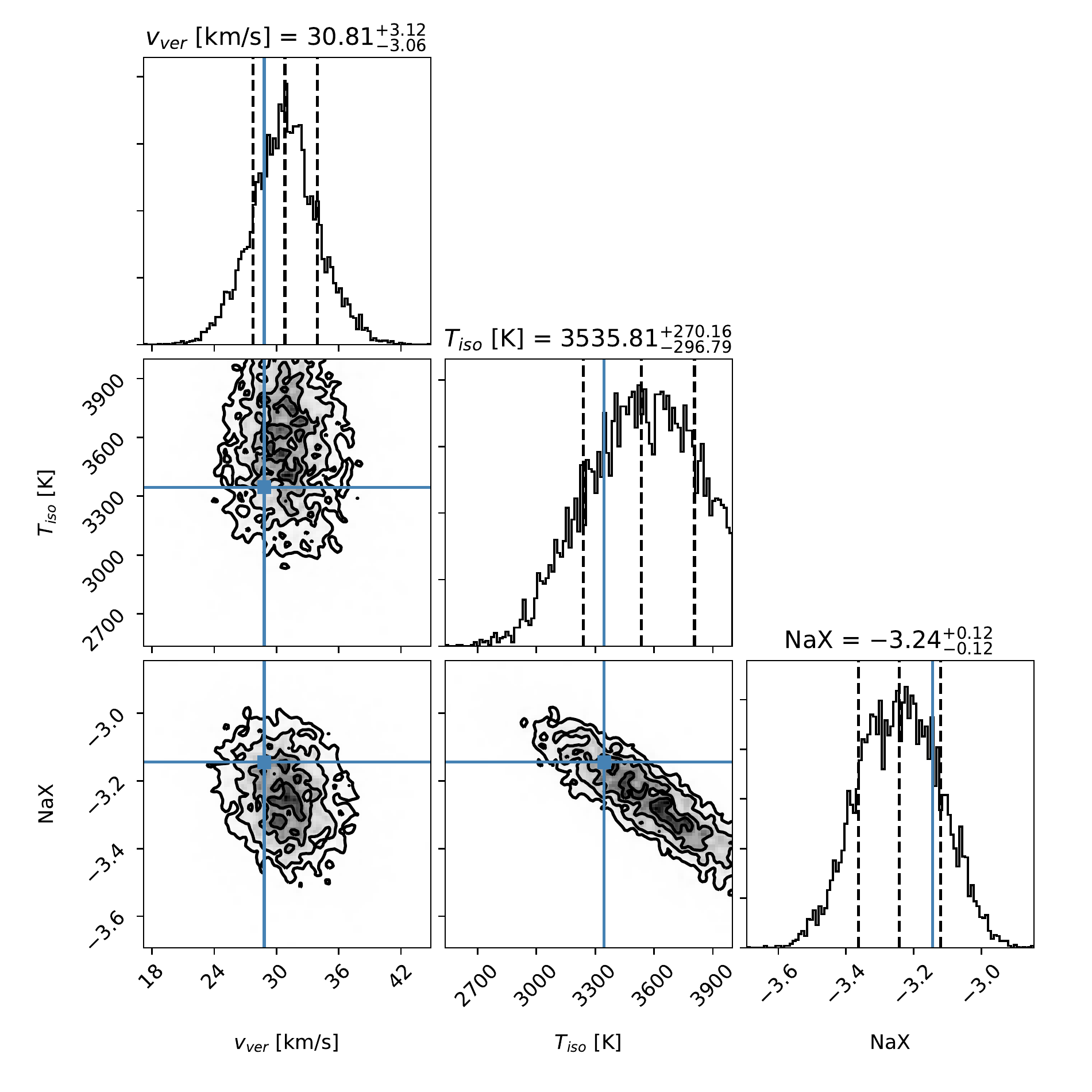}}
        \caption{Posterior distribution of isothermal line retrieval with an added vertical wind constant throughout the atmosphere.}
        \label{fig:vverticalposterior}
\end{figure}

\begin{figure}[htb!]
\resizebox{\columnwidth}{!}{\includegraphics[trim=0.0cm 0.0cm 0.0cm 0.0cm]{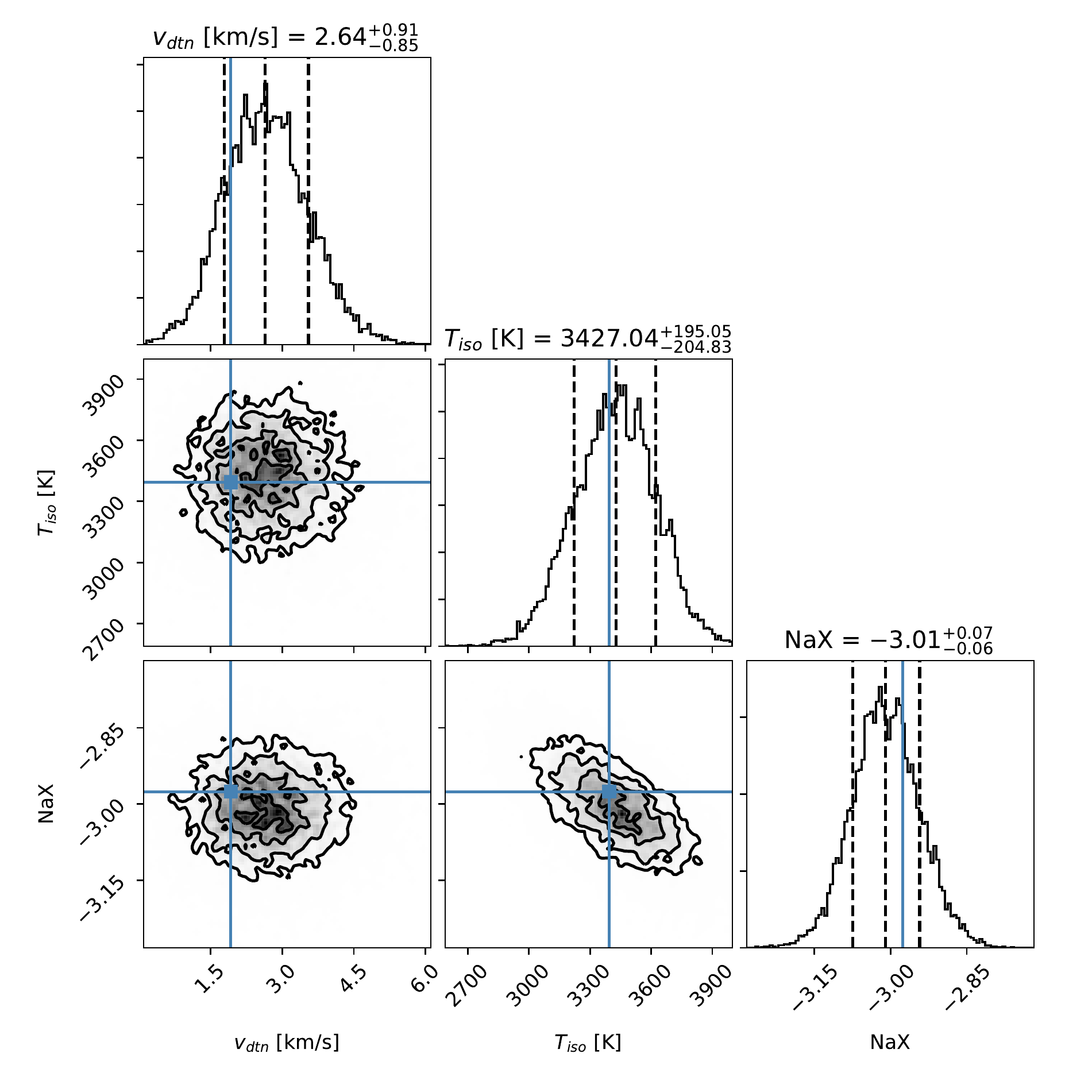}}
        \caption{Posterior distribution of isothermal line retrieval with an added day-to-night side wind constant throughout the atmosphere. }
        \label{fig:dtnposterior}
\end{figure}
\begin{figure}[htb!]
\resizebox{\columnwidth}{!}{\includegraphics[trim=0.0cm 0.0cm 0.0cm 0.0cm]{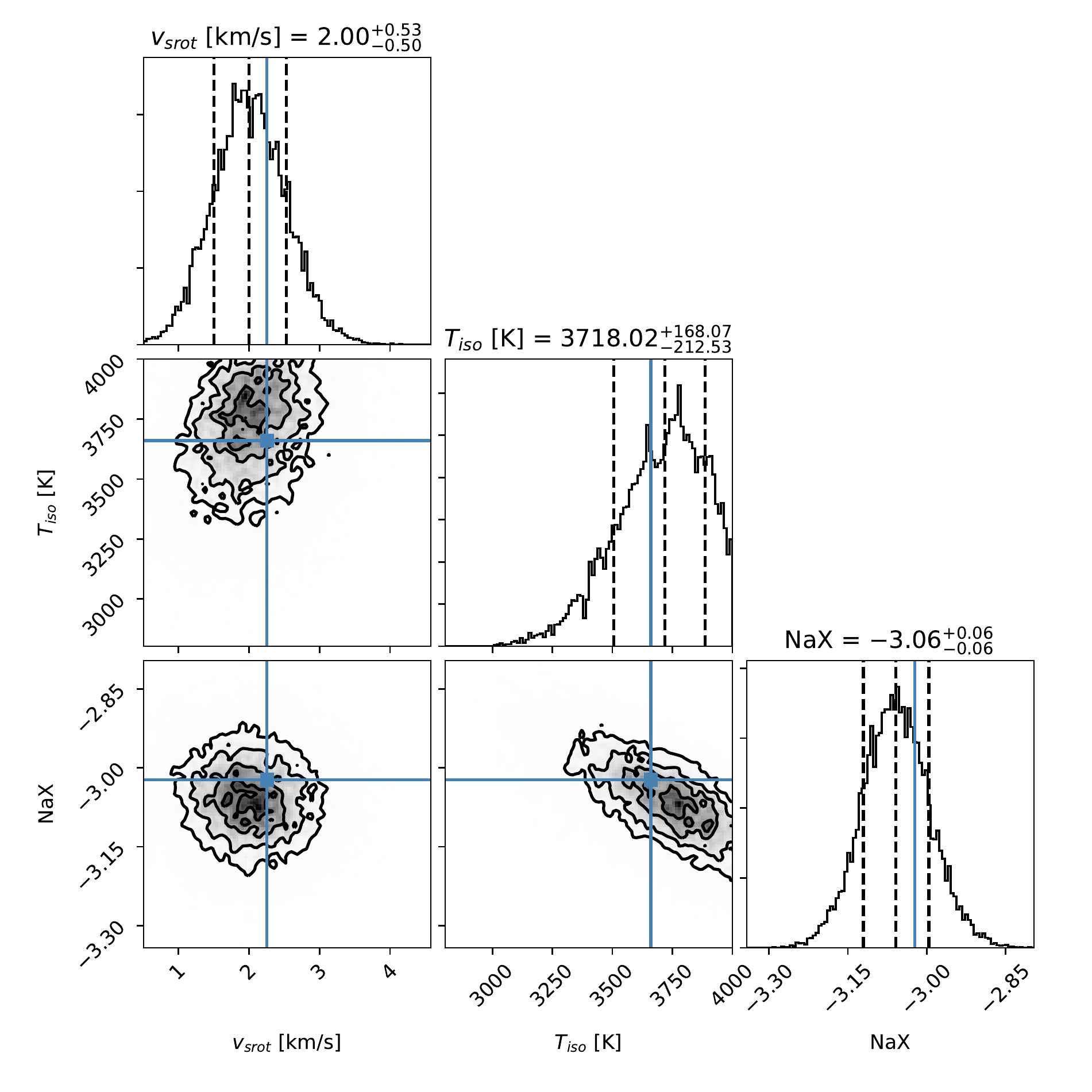}}
        \caption{Posterior distribution of isothermal line retrieval with an added super-rotational wind constant throughout the atmosphere. }
        \label{fig:superrotposterior}
\end{figure}

\begin{figure}[htb!]
\resizebox{\columnwidth}{!}{\includegraphics[trim=0.0cm 0.0cm 0.0cm 0.0cm]{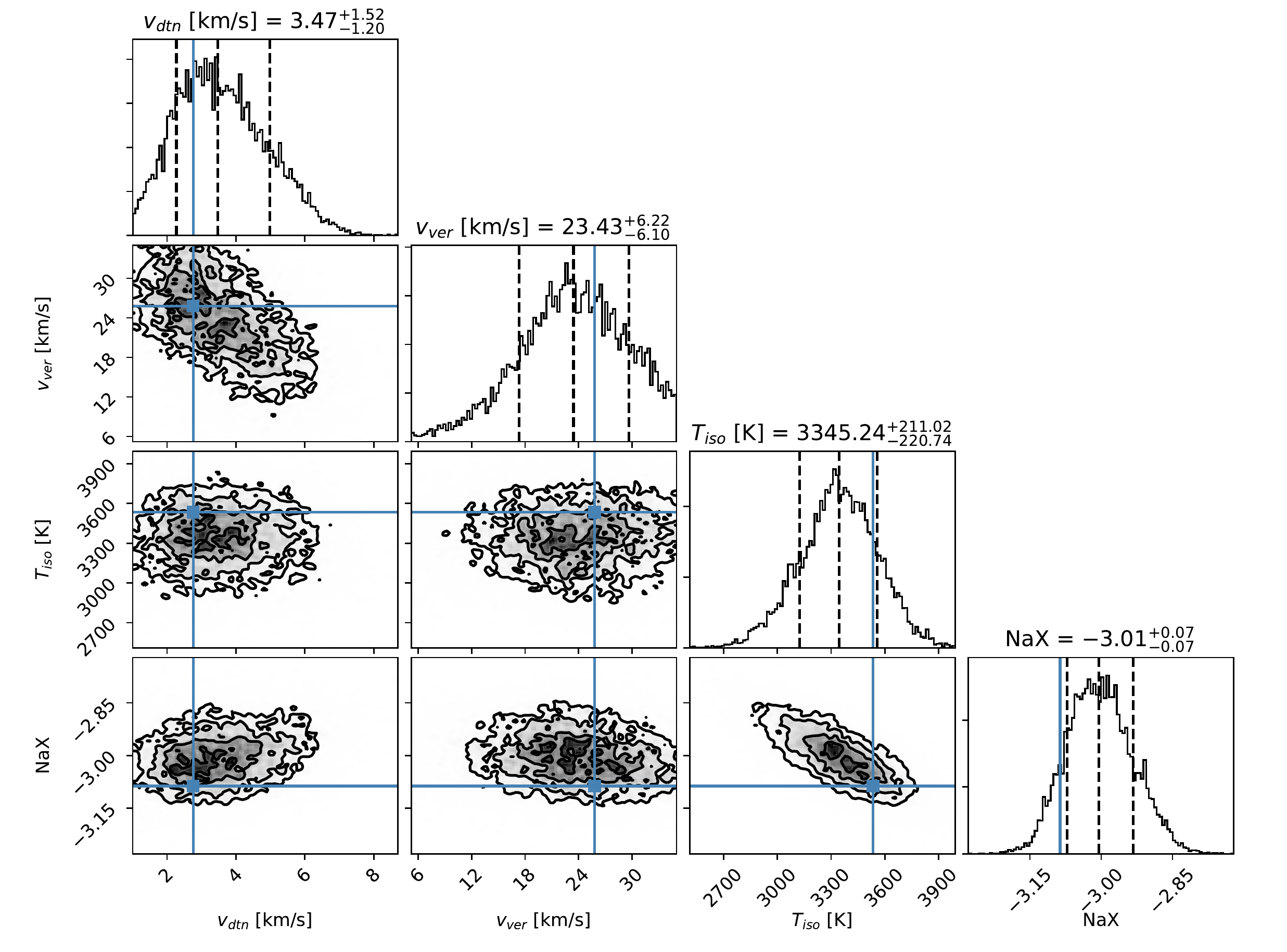}}
        \caption{Posterior distribution of isothermal line retrieval with an added day-to-night side wind in the lower atmosphere and a vertical wind in the upper atmosphere. The change of layers was set at $p=10^{-4}$~bar, with the surface pressure at $P_0=10$~bar.}
        \label{fig:dtn_verbestposterior3}
\end{figure}

\begin{figure}[htb!]
\resizebox{\columnwidth}{!}{\includegraphics[trim=0.0cm 0.0cm 0.0cm 0.0cm]{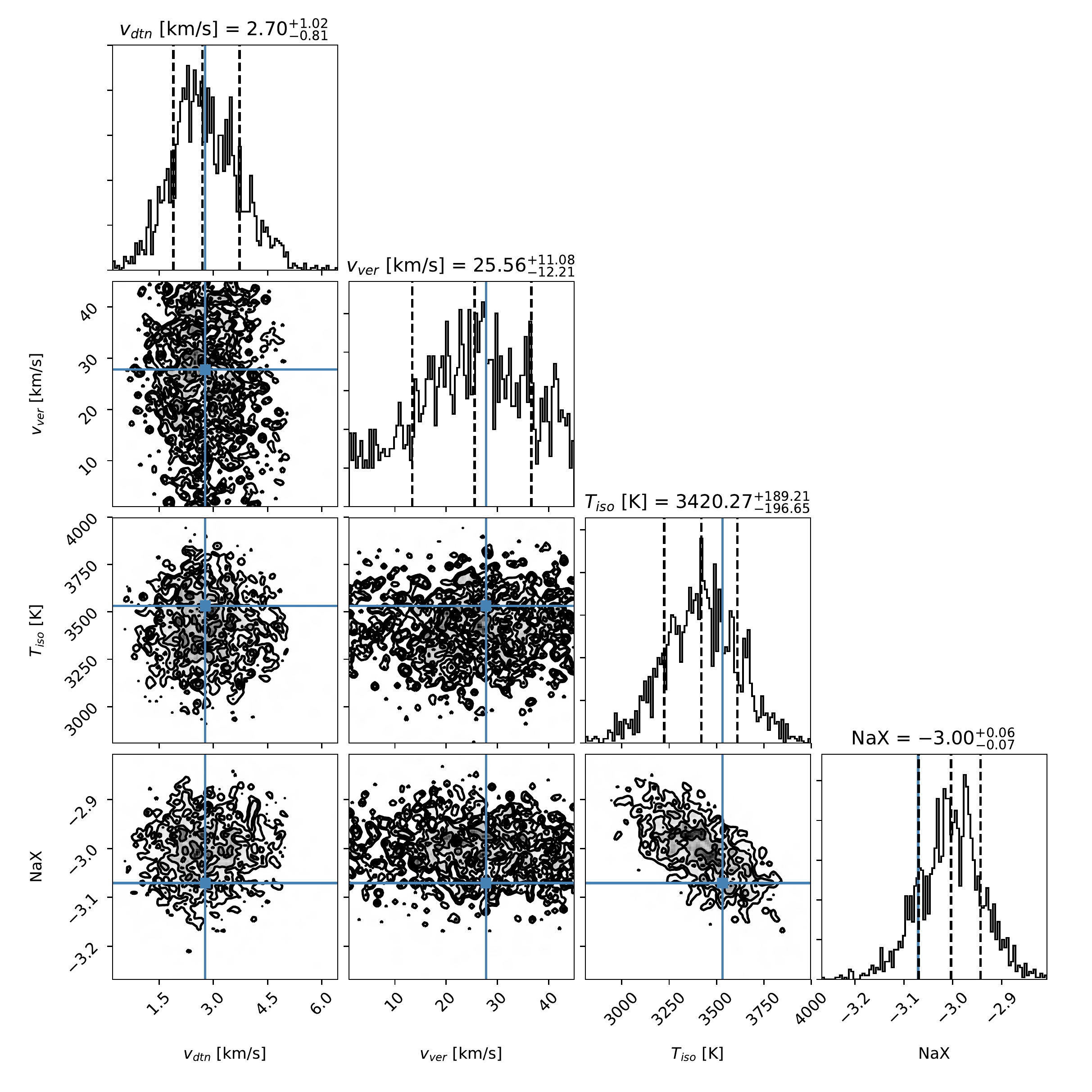}}
        \caption{Posterior distribution of isothermal line retrieval with an added day-to-night side wind in the lower atmosphere and a vertical wind in the upper atmosphere. The change of layers was set at $p=10^{-5}$~bar, with the surface pressure at $P_0=10$~bar.}
        \label{fig:dtn_verbestposterior4}
\end{figure}

\begin{figure}[htb!]
\resizebox{\columnwidth}{!}{\includegraphics[trim=0.0cm 0.0cm 0.0cm 0.0cm]{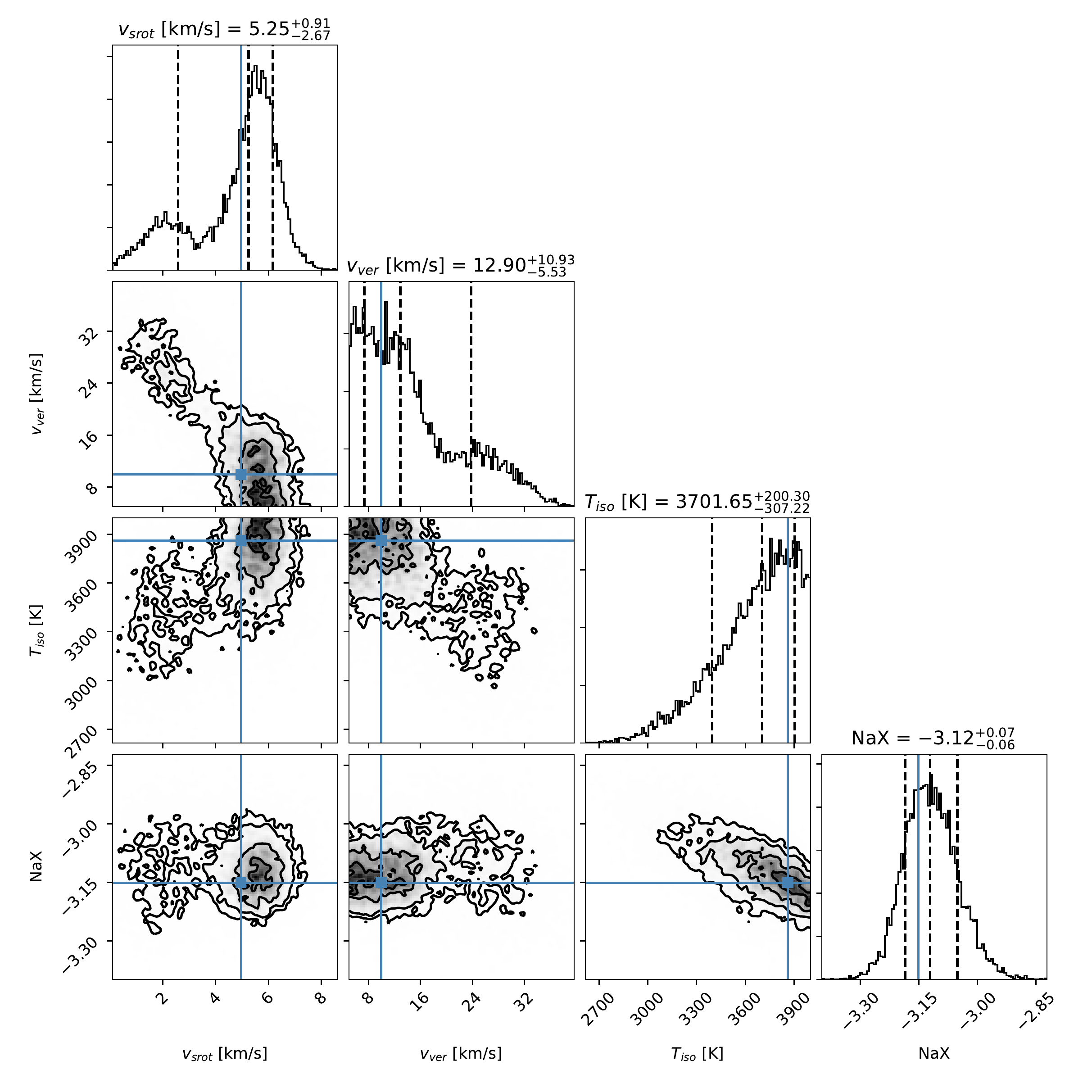}}
        \caption{Posterior distribution of isothermal line retrieval with an added super-rotational side wind in the lower atmosphere and a vertical wind in the upper atmosphere. The change of layers was set at $p=10^{-3}$~bar, with the surface pressure at $P_0=10$~bar.}
        \label{fig:srot_ver2}
\end{figure}

\begin{figure}[htb!]
\resizebox{\columnwidth}{!}{\includegraphics[trim=0.0cm 0.0cm 0.0cm 0.0cm]{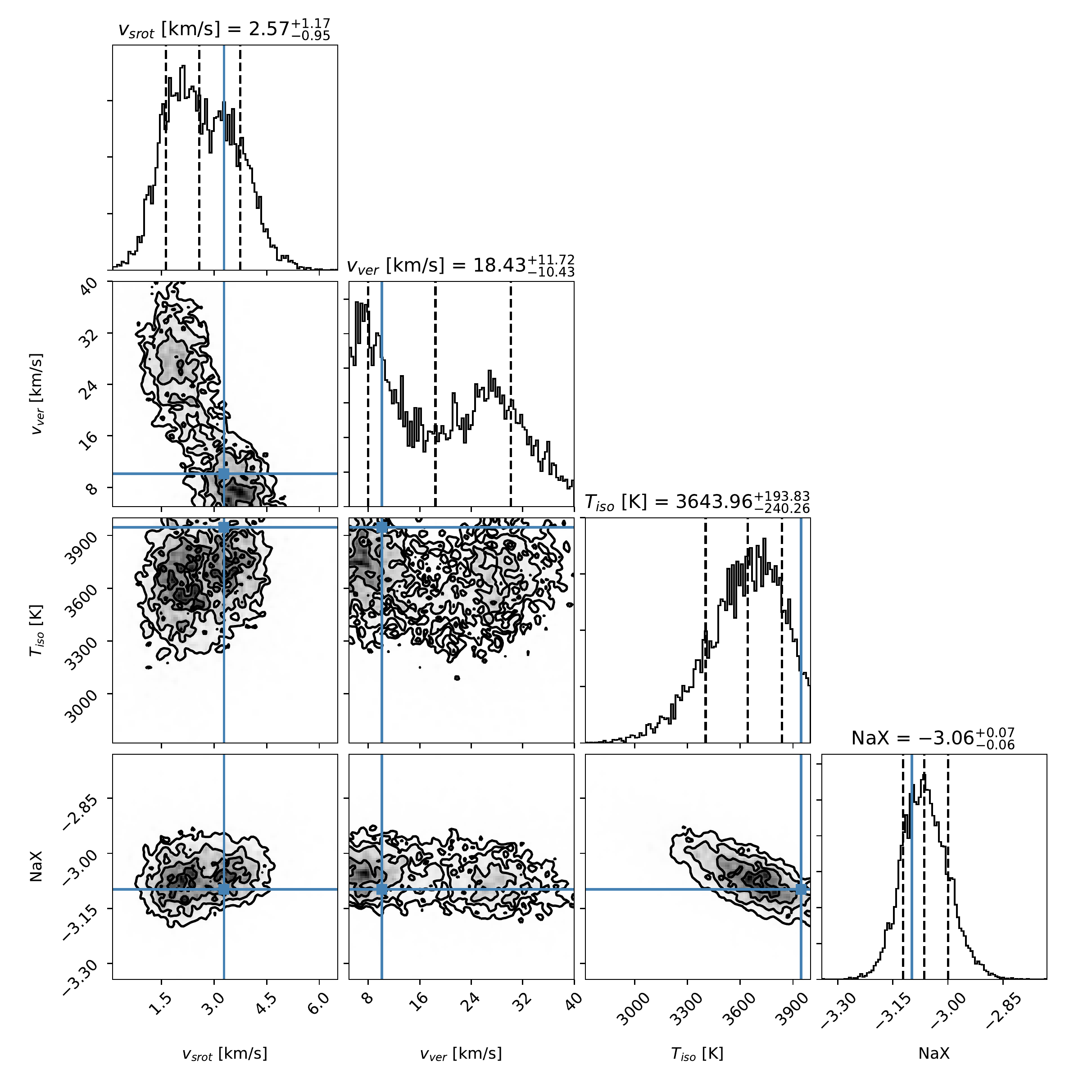}}
        \caption{Posterior distribution of isothermal line retrieval with an added super-rotational side wind in the lower atmosphere and a vertical wind in the upper atmosphere. The change of layers was set at $p=10^{-4}$~bar, with the surface pressure at $P_0=10$~bar.}
        \label{fig:srot_ver3}
\end{figure}

\begin{figure}[htb!]
\resizebox{\columnwidth}{!}{\includegraphics[trim=0.0cm 0.0cm 0.0cm 0.0cm]{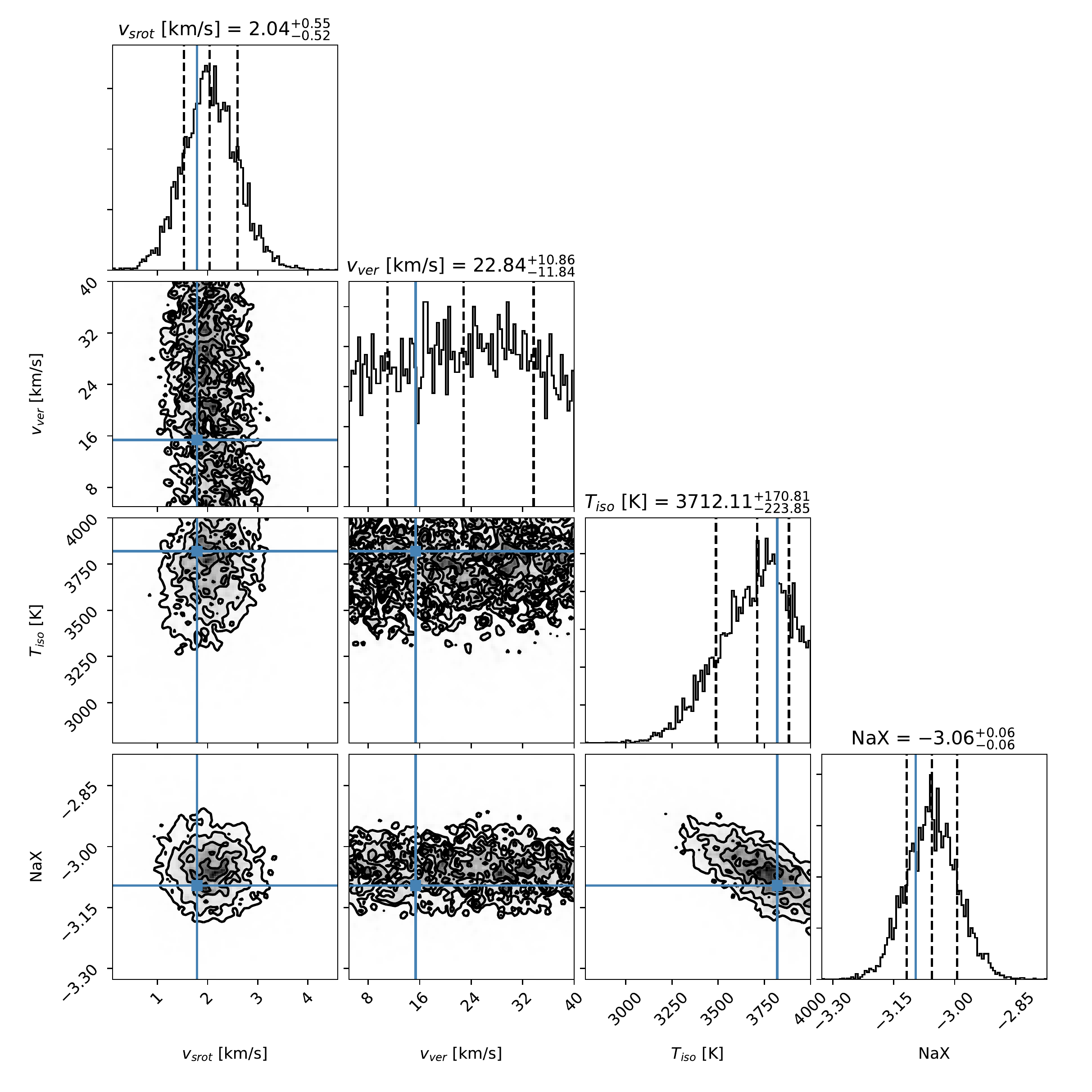}}
        \caption{Posterior distribution of isothermal line retrieval with an added super-rotational side wind in the lower atmosphere and a vertical wind in the upper atmosphere. The change of layers was set at $p=10^{-5}$~bar, with the surface pressure at $P_0=10$~bar.}
        \label{fig:srot_ver4}
\end{figure}
\begin{figure}[htb!]
\resizebox{\columnwidth}{!}{\includegraphics[trim=0.0cm 0.0cm 0.0cm 0.0cm]{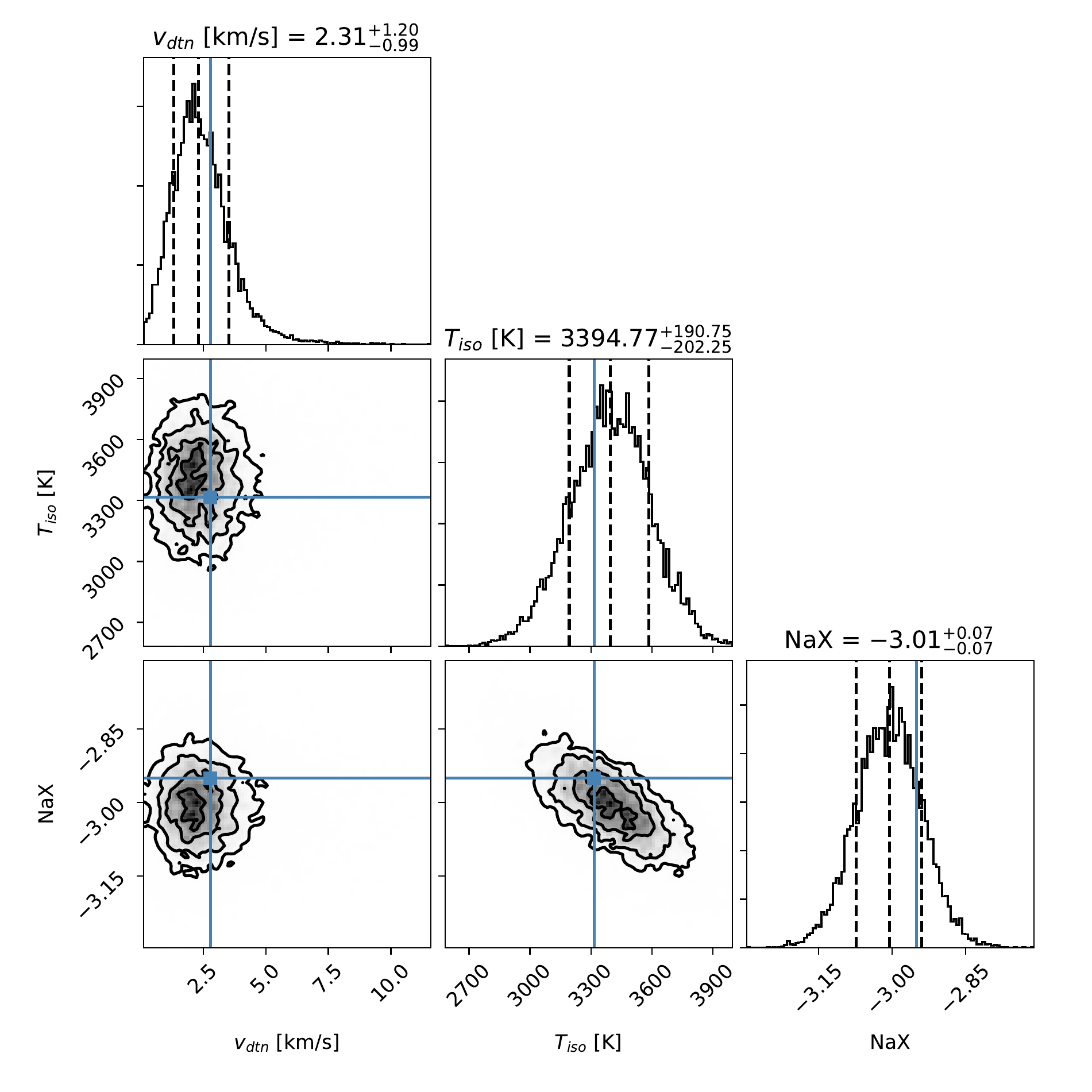}}
        \caption{Posterior distribution of isothermal line retrieval with an added day-to-night side wind dependent on the latitude via $\cos\theta$. }
        \label{fig:vdtnposteriorcos}
\end{figure}
\begin{figure}[htb!]
\resizebox{\columnwidth}{!}{\includegraphics[trim=0.0cm 0.0cm 0.0cm 0.0cm]{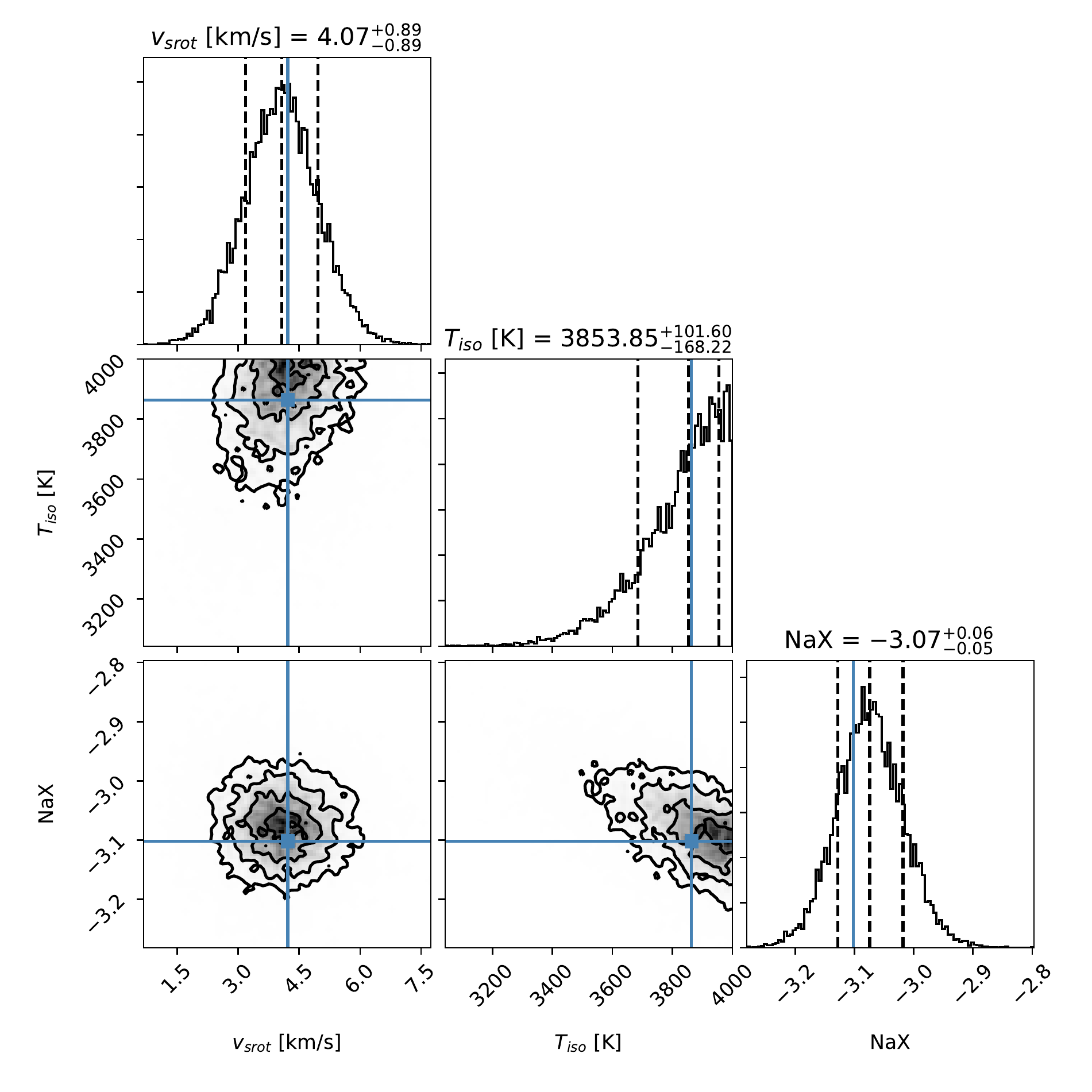}}
        \caption{Posterior distribution of isothermal line retrieval with an added super-rotational wind dependent on the latitude via $\cos\theta$. }
        \label{fig:superrotposteriorcos}
\end{figure}

\begin{figure}[htb!]
\resizebox{\columnwidth}{!}{\includegraphics[trim=0.0cm 0.0cm 0.0cm 0.0cm]{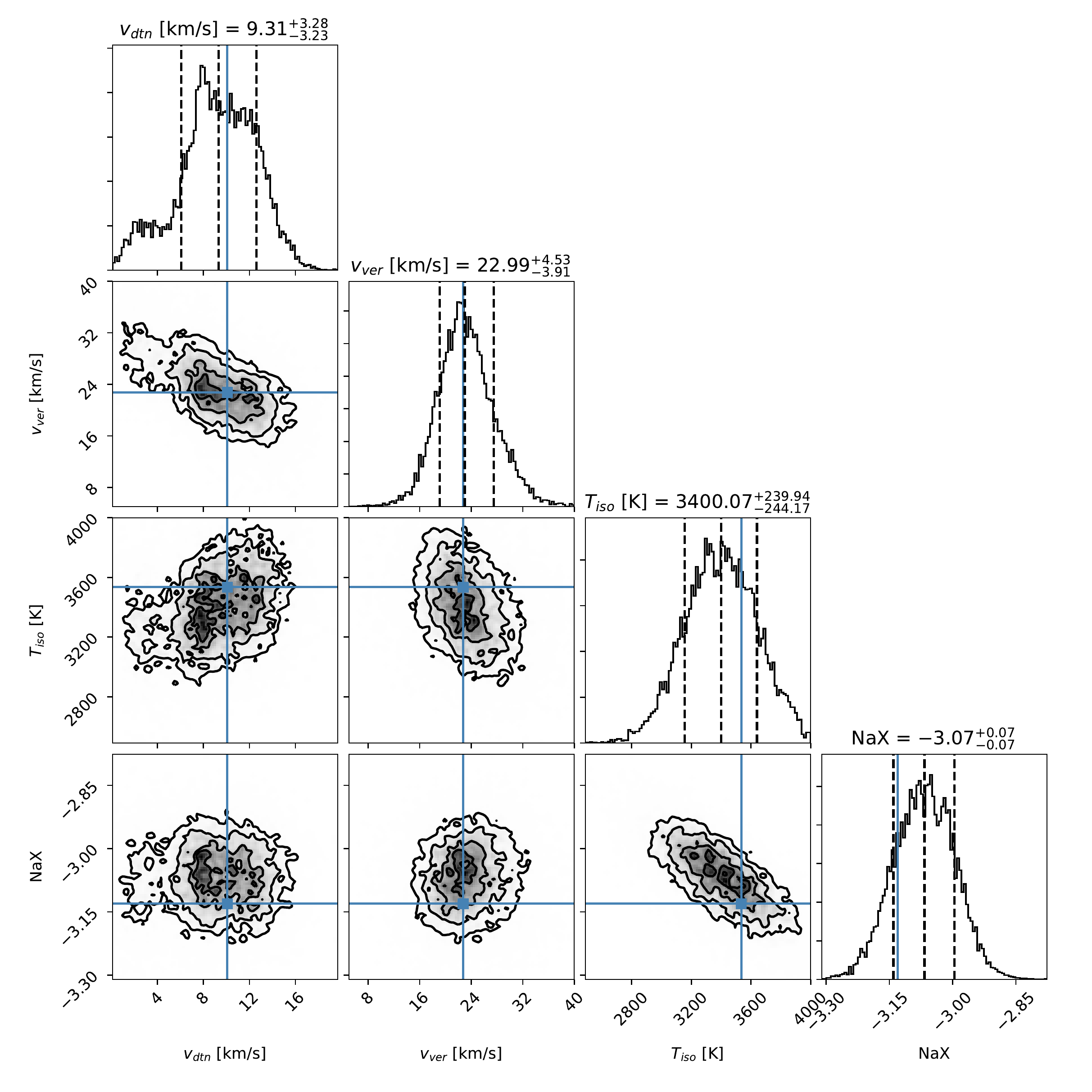}}
        \caption{Posterior distribution of isothermal line retrieval with an added $\cos\theta$ dependent day-to-night side wind in the lower atmosphere and a vertical wind in the upper atmosphere. The change of layers was set at $p=10^{-3}$~bar, with the surface pressure at $P_0=10$~bar.}
        \label{fig:dtn_verposteriorcos}
\end{figure}


\FloatBarrier
\section{Discussion on the planetary orbital velocity}
\label{app:kpvsys}
\begin{figure}[htb!]
\resizebox{\columnwidth}{!}{\includegraphics[trim=3.0cm 8.0cm 3.0cm 9.0cm]{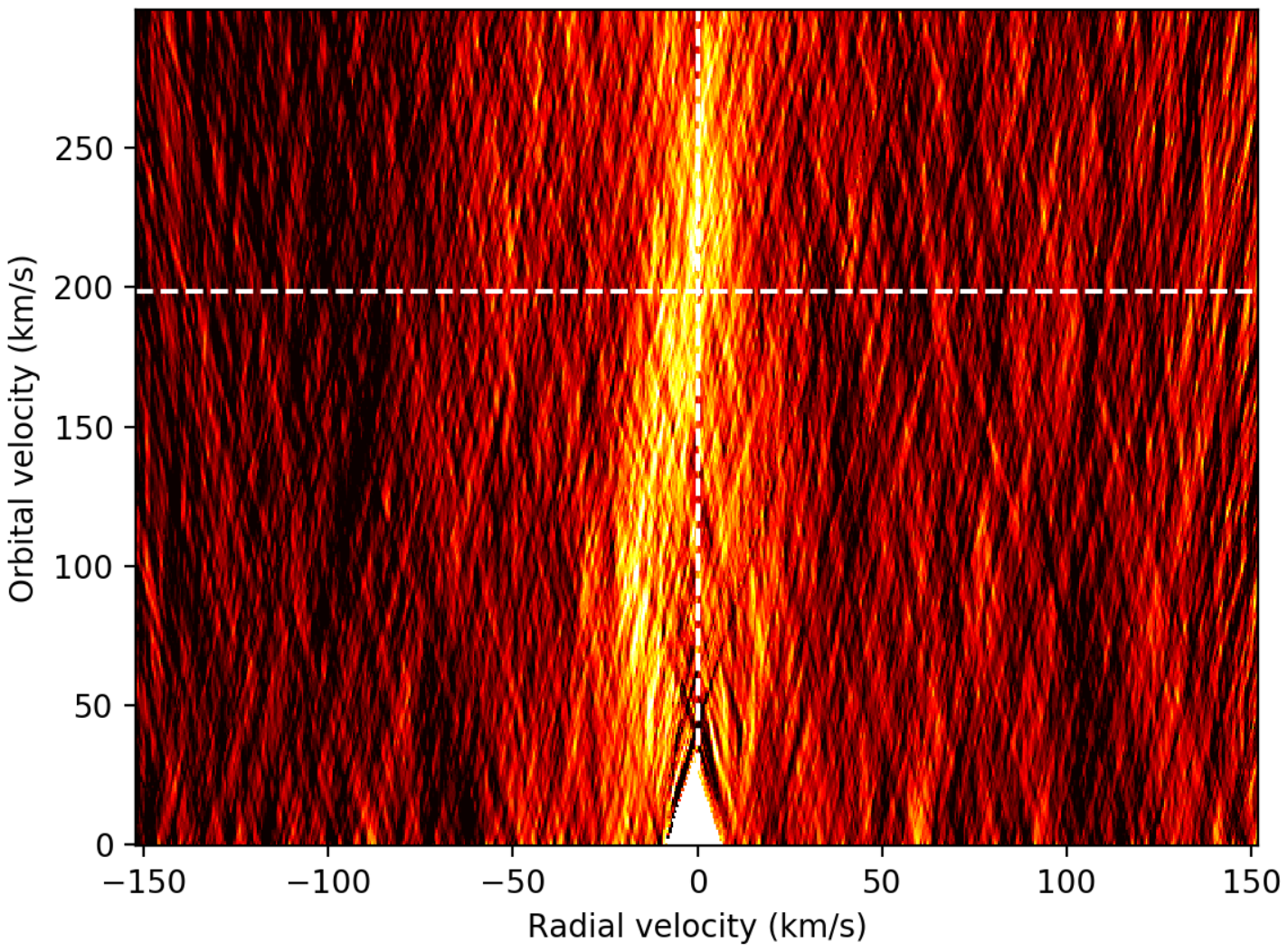}}
        \caption{$K_p$-$v_{\mathrm{sys}}$ map generated from the two ESPRESSO nights, the position of the sodium line as calculated in \citet{Ehrenreich2020} is indicated as white, dashed lines. White areas stem from the masking of the data at the centre of the stellar sodium lines. Lighter colours indicate a stronger detection signal.}
        \label{fig:KpVsys}
\end{figure}

\noindent To strengthen the here presented argument that the line shape stems indeed from the dynamical structure of the exoplanet atmosphere and not from factors related to the orbital motion, we have studied the impact of the planetary orbital motion ($K_p$) and its uncertainty on our results. In our analysis, we have used the value of $K_p$ as calculated from the stellar mass, period and inclination in \citet{Ehrenreich2020}. 

However, to demonstrate the impact of $K_p$ on the sodium doublet's line shape, we have also derived $K_p$ directly from the signal of the planetary sodium (see Figure \ref{fig:KpVsys}). Figure \ref{fig:KpVsys} shows the $K_p$ - $v_{\mathrm{sys}}$ space, where we have combined both ESPRESSO transits and both sodium lines to generate the signal. We have excluded the HARPS data for this demonstration to keep rebinning at a minimum and the signal shape as precise as possible. 

\noindent The map presented in Figure \ref{fig:KpVsys} confirms the sodium signal already independently established with the HARPS spectrograph \citep{Seidel2019} and the ESPRESSO spectrograph \citep{Tabernero2020} and also confirms its broad line shape. However, the map demonstrates as well the difficulties encountered when deriving $K_p$ directly from the data of single lines instead of hundreds or even thousands of spectral lines (see e.g. \citealt{Hoeijmakers2020}). The signal is too broad in $K_p$ to properly derive a precise value, spanning from roughly $50$ to $250\,\kms$ due to the low number of lines available.\\
 
It also demonstrates that the line shape is independent from small errors on $K_p$, since we sample very similarly broad profiles for values of $K_p$ varying up to $10\,\kms$ from the value derived in \citet{Ehrenreich2020}. The value of $K_p$ as calculated in \citet{Ehrenreich2020} is, therefore, preferred to our own calculation for two reasons: it is much more precise with a value of $196.52\pm 0.94\,\kms$, which translates to a maximum error at ingress and egress of $\pm 0.3\,\kms$ (less than one ESPRESSO pixel). Additionally, in their calculation, they do not take the in-transit data into account, decoupling the calculation of $K_p$ from any atmospheric dynamics. This makes any retrieval of the atmospheric dynamics more robust against the influence of orbital motion.

In conclusion, the derivation of $K_p$ from our dataset has shown that the sodium signal exists and that its line shape can only be explained by broadening winds or other atmospheric dynamics and that the influence of the orbital motion on our results is negligible.


\end{document}